\documentclass[fleqn,usenatbib]{mnras}

\usepackage[T1]{fontenc}

\DeclareRobustCommand{\VAN}[3]{#2}
\let\VANthebibliography\thebibliography
\def\thebibliography{\DeclareRobustCommand{\VAN}[3]{##3}\VANthebibliography}

\usepackage{graphicx}	
\usepackage{amsmath}	
\usepackage{amssymb}	
\usepackage[labelfont=bf]{subcaption} 
\usepackage{booktabs} 
\usepackage{threeparttable} 

\graphicspath{ {graphics/} } 
\defcitealias{Caldwell98}{C98}

\title[A Tidally-Disrupting Ultra-Diffuse Galaxy]{A Tale of a Tail: A Tidally-Disrupting Ultra-Diffuse Galaxy in the M81 Group}

\author[Rokas \v{Z}emaitis et al.]{\href{https://orcid.org/0000-0002-8566-0491}{Rokas \v{Z}emaitis$^{1}$\thanks{E-mail: rokasz@roe.ac.uk (RZ)}},
\href{https://orcid.org/0000-0001-7934-1278}{Annette M. N. Ferguson$^{1}$},
\href{https://orcid.org/0000-0002-7866-0514}{Sakurako Okamoto$^{2,3,4}$},
\href{https://orcid.org/0000-0002-3263-8645}{Jean-Charles Cuillandre$^{5}$},
\newauthor
\href{https://orcid.org/0000-0002-9086-6398}{Connor J. Stone$^{6}$}, 
Nobuo Arimoto$^{3,7}$,
\href{https://orcid.org/0000-0002-2191-9038}{Mike J. Irwin$^{8}$}
\\
$^{1}$Institute for Astronomy University of Edinburgh, Blackford Hill, Edinburgh EH9 3HJ UK\\
$^{2}$Subaru Telescope, National Astronomical Observatory of Japan, 650 North A'ohoku Place, Hilo, HI 96720, USA\\
$^{3}$National Astronomical Observatory of Japan, Osawa 2-21-1, Mitaka, Tokyo, 181-8588, Japan\\
$^{4}$The Graduate University for Advanced Studies, Osawa 2-21-1, Mitaka, Tokyo 181-8588, Japan\\
$^{5}$Universit\'{e} Paris-Saclay, Universit\'{e} Paris-Cit\'{e}, CEA, CNRS, AIM, 91191, Gif-sur-Yvette, France\\
$^{6}$ Department of Physics, Engineering Physics \& Astronomy, Queen's University, Kingston, ON K7L 3N6, Canada\\
$^{7}$Astronomy Program, Department of Physics and Astronomy, Seoul National University, 599 Gwanak-ro, Gwanak-gu, Seoul 151-742, Korea\\
$^{8}$Institute of Astronomy, Madingley Road, Cambridge CB3 0HA UK
}

\date{Accepted XXX. Received YYY; in original form ZZZ}

\pubyear{2022}

\begin{document}
\label{firstpage}
\pagerange{\pageref{firstpage}--\pageref{lastpage}}
\maketitle

\begin{abstract}
We present the discovery of a giant tidal tail of stars associated with F8D1,  the closest known example of an ultra-diffuse galaxy (UDG). F8D1 sits in a region of the sky heavily contaminated by Galactic cirrus and has been poorly studied since its discovery two decades ago.   The tidal feature was revealed in a deep map of resolved red giant branch stars constructed using data from our Subaru Hyper Suprime-Cam survey of the M81 Group.  It has an average surface brightness of $\mu_g \sim 32$ mag arcsec$^{-2}$  and can be traced for over a degree on the sky (60~kpc at the distance of F8D1) with our current imagery.  We revisit the main body properties of F8D1 using deep multiband imagery acquired with MegaCam on CFHT and measure effective radii of 1.7-1.9~kpc, central surface brightnesses of $24.7-25.7$ mag and a stellar mass of $\sim7 \times 10^7 M_{\odot}$.  Assuming a symmetric feature on the other side of the galaxy, we calculate that $30-36$\% of F8D1's present-day luminosity is contained in the tail.  We argue that the most likely origin of F8D1's disruption is a recent close passage to M81, which would have stripped its gas and quenched its star formation.  As the only UDG that has so far been studied to such faint surface brightness depths, the unveiling of F8D1 ‘s tidal disruption is important.  It leaves open the possibility that many other UDGs could be the result of similar processes, with the most telling signatures of this lurking below current detection limits.
\end{abstract}

\begin{keywords}
galaxies: individual: F8D1 -- galaxies: interactions -- galaxies: groups: individual: M81 Group -- galaxies: structure -- galaxies: stellar content
\end{keywords}



\section{Introduction}

Ultra-diffuse galaxies (UDGs) are distinguished by their low central surface brightnesses and large sizes.  A commonly-used, albeit rather arbitrary, definition is that they have effective radii $R_{\rm eff} \geq $ 1.5 kpc and central surface brightnesses fainter than $\mu_{g}(0) \gtrsim 24$ mag~arcsec$^{-2}$ \citep[e.g.,][]{vanDokkum2015,Koda2015}. Although systems with these properties have been known to exist for decades \citep[e.g.,][]{Sandage1984, Caldwell1987, Impey1988}, there has been a recent resurgence in interest in them due to the sheer abundance of such objects being discovered in modern deep imaging surveys.  They appear particularly common in dense environments \citep[e.g.,][]{vanDokkum2015,Koda2015, Mihos2015, Janssens2017, Iodice2020}, but are also found in low density groups and in the field \citep[e.g.,][]{Merritt2016, Roman2017, Greco2018, Muller2018}. 

Studies of UDGs in the last few years have yielded important new information about the properties of these unusual galaxies. It has been established that many UDGs have round isophotes and radial distributions of starlight that follow an exponentially-declining or flatter profile \citep[e.g.,][]{Yagi2016,Cohen2018, Alabi2020}. Their star formation properties are observed to vary, with systems which ceased to form stars a long time ago dominating in cluster environments while the field contains examples where  star formation has continued until recent epochs  \citep[e.g.,][]{Leisman2017,Greco2018,Prole2019}. 

Much debate has centered on the dark matter content of UDGs and, again, this seems to be a property with considerable variance. There have been extremely dark matter-dominated UDGs reported as well as systems that appear to have very little dark matter at all  \citep[e.g.,][]{Toloba2018, Lim2018, Danieli2019, Forbes2020}. A common way to infer the amount of dark matter in a UDG is to exploit the relationship between the number of globular clusters (GCs) in a system and the host halo mass \citep[e.g.,][]{Spitler2009,Harris2013,Forbes2018}.  While this is in principle a simple technique, results can vary considerably depending on how it is applied in practice. For example, while \citet{vanDokkum2017} find the Coma cluster UDG DF44 to have $\sim75$ GCs associated with it, \cite{Saifollahi2021} use the same dataset to argue that this number should instead be $\sim21$.  Other studies support the idea that the GC systems of UDGs are generally consistent with them being dwarf galaxies with normal dark matter content  \citep[e.g.,][]{Amorisco2018, Lim2018}. 

Several explanations for the origin of UDGs have appeared in the literature. Firstly, they may be rapidly rotating normal dwarf galaxies, resulting in a very  extended profile for their stellar mass \citep{Amorisco2016}. Alternately, their large sizes may result from tidal stripping and heating experienced through interactions with massive neighbouring galaxies \citep{Carleton2019, Doppel2021} or through formation in tidal debris \citep{Bennet2018}. Other possible scenarios  for the origin of UDGs are that they are puffed-up dwarf galaxies that have experienced gas loss due to star formation-driven outflows \citep{diCintio2017, Jiang2019} or that they are `failed' galaxies that did not manage to build-up their expected stellar mass \citep{vanDokkum2015}. Ram-pressure stripping may play a further role in transforming gas-rich blue UDGs into red ones \citep[e.g.,][]{Junais2021}.  While it is likely that the present-day UDG population results from a variety of these formation channels, the question of which is the dominant channel remains open.  

\begin{figure*}
\centering
\includegraphics[width=\textwidth]{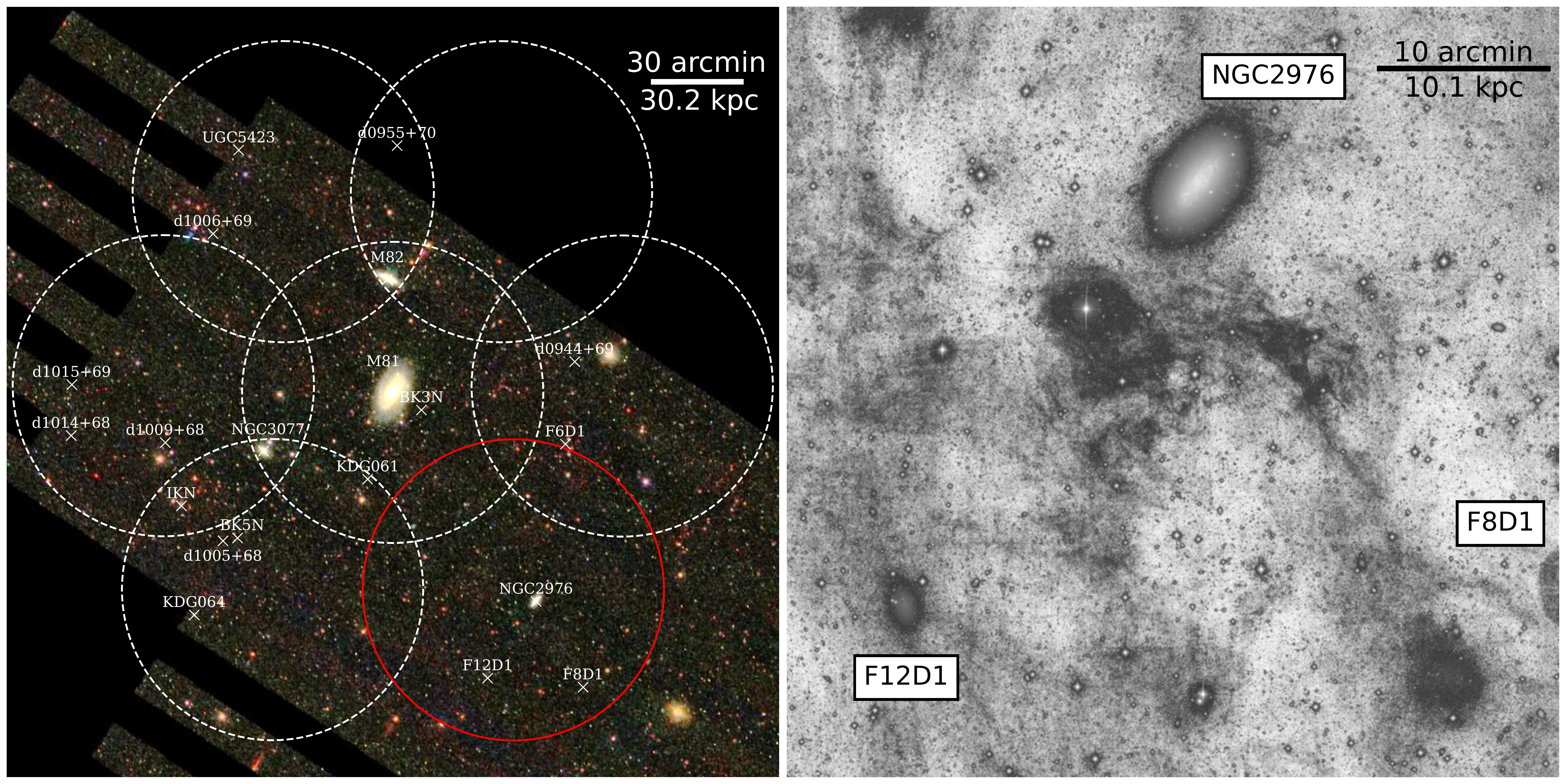}
\caption{Left: Our M81 Group Survey footprint (7 HSC pointings) overlaid on
an SDSS image. Each HSC pointing has a diameter of 1.5 degrees and the known galaxies within this area are marked. The red circle indicates the pointing containing F8D1, which is the focus of this paper. Right: A portion of a deep {\it i}-band image taken with CFHT/MegaCam that shows a zoom-in on the central region of the red-circled HSC pointing. Highly-structured  cirrus is present throughout this region of the sky which greatly complicates integrated light studies of the low surface brightness emission of F8D1 and its neighbours. F8D1 appears projected over a particularly bright ISM filament. The scalebar in both panels reflects the physical scale at the distance at M81. North is up and east is to the left.} \label{fig:full_map}
\end{figure*}

Local universe examples of UDGs are particularly valuable since their proximity allows for much greater insight into their properties compared to more distant systems.  Some very low surface brightness extended dwarf galaxies have been uncovered in the nearby universe. \citet{Toloba2016} discuss the NGC\,253 satellite Scl-MM-Dw2 which has $R_{\rm eff} = $ 2.9 kpc, $\mu_{V}(0) = 27.7$ mag~arcsec$^{-2}$ and $M_V= -12$, while \citet{Martin2016} find the M31 satellite And\,XIX to have $R_{\rm eff} = $ 3.1 kpc, $\mu_{V}(0) = 29.3$ mag~arcsec$^{-2}$ and $M_V= -10$.  These may be extreme examples of the UDG category but they are not direct analogues of the dominant population of UDGs seen at greater distances. While most famously known for its triplet of interacting galaxies (M81-M82-NGC3077), the  nearby M81 Group (D = 3.6 Mpc; \citealt{Radburn-Smith2011}) hosts a low luminosity galaxy with properties that unequivocally place it in 
the UDG realm.  F8D1 was discovered more than 20 years ago by \citet[][hereafter \citetalias{Caldwell98}]{Caldwell98} during a CCD survey for M81 Group analogues to Local Group dwarf elliptical galaxies. It has a luminosity of $M_V\sim -14$ or $\sim 4\times10^7$ L$_{\odot}$ and lies roughly 2 degrees in projection from M81 in a region that is heavily contaminated by Galactic cirrus \citep{Sandage1976}. Although \citetalias{Caldwell98}  remarked on the fact that its low central surface brightness ($\mu_{V}(0)\sim 25.4$ mag~arcsec$^{-2}$) and large effective radius  ($R_{\rm eff} \sim 2.5$~kpc) were in stark contrast to those of most galaxies known at that time, F8D1 has barely received any attention beyond its initial discovery.  Using {\it Hubble Space Telescope} Wide Field Planetary Camera 2 (HST/WFPC2) data, \citetalias{Caldwell98} were able to resolve the upper two magnitudes of the red giant branch (RGB) of F8D1 and use the luminosity of the RGB tip to firmly place it at the distance of the M81 Group \citep[see also][]{Kara2000}.  They also used the colour of the RGB to measure the mean metallicity of the galaxy as [Fe/H] = $-1.0\pm0.26$ dex, with no significant radial abundance gradient and showed that it hosts a considerable population of luminous asymptotic giant branch (AGB) stars. Such a population is a testament to an extended period of star formation;  
\citetalias{Caldwell98} argued that roughly 30\% of the stellar population was of intermediate age, with the youngest stars having formed only 3-4 Gyr ago. A single globular cluster candidate was identified near the centre of the galaxy. This object was selected on the basis of its semi-resolved morphology and integrated colour. It is rather blue ($V-I\sim0.76$) which could be indicative of a low metallicity. On the other hand, \citet{Chiboucas2009} report a radial velocity of $-125\pm130$~km~s$^{-1}$ for this object which leaves open the possibility it is a background galaxy. Unfortunately, there has been no published radial velocity for F8D1 yet.

In this paper, we revisit the properties of F8D1 -- in retrospect, arguably the closest currently-known {\it bona fide} UDG to the Milky Way\footnote{\citet{Trujillo2017} have previously reported the system UGC 2162, lying at 12.3~Mpc as the closest UDG with a main body that is still largely intact.} -- using new observations and report the discovery of an enormous ($\gtrsim 60$~kpc)  tidal stream emanating from this system.  This provides direct evidence that the unusual properties of F8D1 are a result of significant tidal stripping and also highlights the thus far unappreciated role of this galaxy in the overall dynamical history of the M81 Group. The primary data are drawn from a deep survey of the M81 Group that we have conducted using the  Hyper Suprime-Cam (HSC) wide-field imager on the 8.2m Subaru telescope. Early findings from this survey were presented in \citet{Okamoto2015} and \citet{Okamoto2019}, and results of the full survey, including stellar halo profiles and a new satellite search, will be the subject of forthcoming papers (\v{Z}emaitis et al., in prep; Okamoto et al., in prep). 

The paper is structured as follows: Section 2 describes the observational data used in the paper together with its reduction and corrections. We present the discovery of the tidal tail in Section 3 and characterise its properties as well as those of the main body of F8D1.  We interpret and discuss our results in Section 4, and  summarise our findings in Section 5.

\section{The Data}
\subsection{HSC Observations and Data Reduction}\label{hsc}

 Our primary dataset consists of observations obtained with HSC on the Subaru Telescope. The HSC imager consists of a mosaic of 104 CCDs, sampling a field-of-view (FOV) of 1.76 sq. deg with 0.17 arcsec pixels \citep{Miyazaki2018}.
 Our M81 Group survey covers $\sim 12$ square degrees and reaches to roughly two magnitudes below the tip of the RGB at the distance of M81. The survey footprint is centered on M81 and comprises seven HSC pointings in total, extending to a projected radius of $\sim 130$ kpc from the centre of M81 (see Fig. \ref{fig:full_map}, left panel). The data were obtained in the course of various classical and queue-mode runs during the period 2015--2019. The survey uses the g- and i-band filters to image to depths of {\it g}$\sim27.5$ and {\it i}$\sim26.5$ (5-sigma), with seeing ranging from 0.6$''$ to 1.0$''$. This depth corresponds to more than two magnitudes below the tip of the RGB at the distance of M81.  In this work, we focus on one particular HSC pointing from the survey, the South-West field highlighted in the left panel of Fig. \ref{fig:full_map} which contains F8D1 near its South-Western edge. The total exposure time for this field is  6160s ($28\times220$s) and 6900s ($32\times210$s) in the {\it g}- and {\it i}-filters, respectively. This field also includes two additional M81 Group members, the low mass spiral NGC\,2976 and the dwarf galaxy F12D1 (also known as [KK98] 077).

The HSC data were processed using the native image reduction pipeline \texttt{hscPipe} 8.4 \citep{Bosch2019}, developed as a precursor to the Rubin Observatory data reduction pipeline \citep{Axelrod2010, Juric2017, Ivezic2019}. The images were corrected for bias, dark count and flat-fielding, and the sky was modelled internally and subtracted from each frame. The frames in each filter were then mosaicked and co-added after photometrically and astrometrically calibrating against Pan-STARRS1 \citep{Magnier2013}.  The final photometry is in the HSC filter system and in AB magnitudes. 

We also used \texttt{hscPipe} to perform PSF-fitting photometry on the co-added {\it g-} and {\it i-}band  frames, forcing the photometry on the {\it i-}band at the positions of the g-band (primary) detections. Experimentation revealed that source detection across the field was optimised when an aggressive 32-pixel (5.4$''$) mesh was used to subtract the sky from each frame.  While this removed the local sky very well and aided in point source detection, it 
rendered the co-adds unusable for diffuse light analyses. The source catalogue was initially cleaned by retaining only sources that have signal-to-noise ratio (SNR) $\geq$ 5 in both bands. Extended sources were subsequently excluded by considering the PSF to cmodel flux ratio.  Following \citet{Pucha2019}, we classify  an object as point-like if the $f_{PSF}/f_{cmodel}$ ratio is within $1\sigma$ of unity at the respective magnitude.

\subsection{Corrections to the Stellar Catalogue}
To assess the completeness of our HSC catalogue, we conduct comparisons to archival HST pointings that fall within our HSC FOV.  We select three Advanced Camera for Surveys (ACS) pointings from GO 16191\footnote{The particular datasets used are: JEEA32010, JEEA32020, JEEA33010, JEEA33020, JEEA35010, JEEA35010.} which consist of exposures of 1103s and 1153s in the F606W and F814W passbands, respectively.  These HST/ACS datasets detect stars to roughly three magnitudes below the TRGB and hence are about a magnitude deeper than our HSC dataset.  As a result, they can be considered to be 100\% complete relative to our ground-based RGB photometry. 

The chosen HST/ACS fields sample a distance range of 6-14 arcmin from the centre of F8D1. Two of the fields lie to the North-East of F8D1, along the major axis and are  roughly coincident with the feature we discuss in Sec. \ref{sec:stream}, whereas the third field lies South-East of the centre along the minor axis. After conducting photometry on the {\it flc} images using DOLPHOT \citep{Dolphin2016}, we searched  for positional matches between stellar sources in the HST/ACS catalogue \citep[selected using the criteria defined in][]{Radburn-Smith2011} and those in our HSC catalogue. We then convert the HST/ACS F606W and F814W magnitudes of the sources onto the HSC {\it g-} and {\it i-}band filter system using transformations that we derived.  Specifically,  we followed the procedure outlined in \cite{Komiyama2018} in which the HSC filter transmission curves were convolved with stellar spectra from the \cite{Gunn1983} and \cite{Pickles1998} spectral atlases. This allowed us to extract synthetic photometry and derive transformations between the HSC and HST  systems across a broad range of colour (see Appendix \ref{sec:transformation} for more detail). Finally, we merge the catalogues of stars in the three HST/ACS fields into one, thus ensuring that the completeness curves do not suffer from statistical noise in less populated magnitude bins.

We bin the stellar sources in varying {\it g}-- and {\it i}--band magnitudes separately, both within the colour limit of 1.0 < $({\it g-i})_0$ < 3.0 to encompass the RGB stars. For each bin, we determine the fraction of the HSC sources which appear in the HST catalogue and characterise the behaviour by fitting Equation 7 of \citet{Martin2016} to the recovered fraction as a function of magnitude.  We find that the {\it i}-band catalogue reaches the 50\% completeness limit at 25.35 mag, while the {\it g}-band catalogue is 50\% complete until 27.06 mag. We use these completeness curves to produce a 2-D histogram that represents the completeness fraction as a function of magnitude and colour and in the subsequent analyses we  correct our photometric catalogue where appropriate. It is worth noting that all the analyses in this paper concern RGB stars above the 50\% completeness level, so these corrections are never dominant. 

The M81 Group lies in a part of the sky with significant and highly variable extinction from foreground dust. In Fig. \ref{fig:extinction}, we show the magnitude of foreground extinction in the {\it g-}band towards the F8D1 field as derived from the \citet{Schlegel1998} reddening map with corrections from \citet{Schlafly2011} and the HSC filter coefficients provided in \citet{Rodrigo2020}.  The values of $A_g$ reach up to 0.6 mag in some regions and so it was necessary to do a star-by-star extinction correction.  We use the extinction-corrected PSF magnitudes for the remainder of this paper, which will be referred to as \textit{g} and \textit{i} hereafter.

\subsection{Ancillary Low Surface Brightness Data}\label{cfht}

As previously mentioned, the sky subtraction method we have adopted within \texttt{hscPipe} optimises the detection of point sources but removes large portions of diffuse galaxy light. We found that reprocessing the data using a global sky subtraction (which corresponds to super-sky pixels of $1000 \times 1000$ pixels in \texttt{hscPipe}) still results in a visible oversubtraction of the galaxy light.  To facilitate integrated light analysis of the main body of F8D1, we therefore acquired deep images in the {\it g,r,i} filters with the 1.1 square degree MegaCam imager on the 3.6m Canada-France-Hawaii Telescope (CFHT) between Feb 6 and Mar 3 2022. The images were obtained using an observing technique that is optimised for low surface brightness (LSB) surveys at CFHT \citep[e.g.,][]{Ferrarese2012,Duc2015}. Using a 7-position large dither captured in a contiguous time window, with relative offsets greater in scale than most extended features encountered in the field such as galaxies (10$'$), a median correction map is derived. This smoothed map with features starting at a scale of $\sim5'$ integrates all systematic illumination effects and restores, upon a properly scaled subtraction, the true night sky background in each individual science image which is a flat pedestal upon which we now see undisturbed compact and extended diffuse sources. Stacks at full depth are obtained by first removing the specific true sky pedestal from each image and scaling the astronomical flux according to a zero point adjustment (accounting for variable atmospheric absorption, etc). Magnitudes are AB in the MegaCam system \citep{Regnault2009}. Two sets of $g$-band exposures (7$\times$300s each, total 4200s), 3 sets of $r$-band exposures (7$\times$300s each, total 6300s), and 1 set of $i$-band exposures (7$\times$300s, total 2100s each) were acquired. The sky was dark to gray and photometric throughout but the seeing was poor (between 1.1 and 1.5$''$) although adequate for diffuse emission. The final science stacks are produced with a scale of  0.561$''$/pixel, still sampling the seeing adequately while boosting the SNR in the diffuse emission. The sky brightness in the $g,r,i$ bands is respectively 21.56, 21.25, 20.15 magnitude per square arcsecond. The limiting surface brightness is usually derived by measuring the residual flatness in the sky background, commensurate with the faintest measurable contrast from a diffuse emission against the flat sky. This is however impossible here due to the prevalence of the bright variable cirrus all over the field. Typical MegaCam LSB performance however adds 7 magnitudes with respect to the sky background, {\it i.e.} $\sim$28.56 magnitude per square arcsecond in the $g$-band.

\begin{figure}
	\includegraphics[width=\columnwidth]{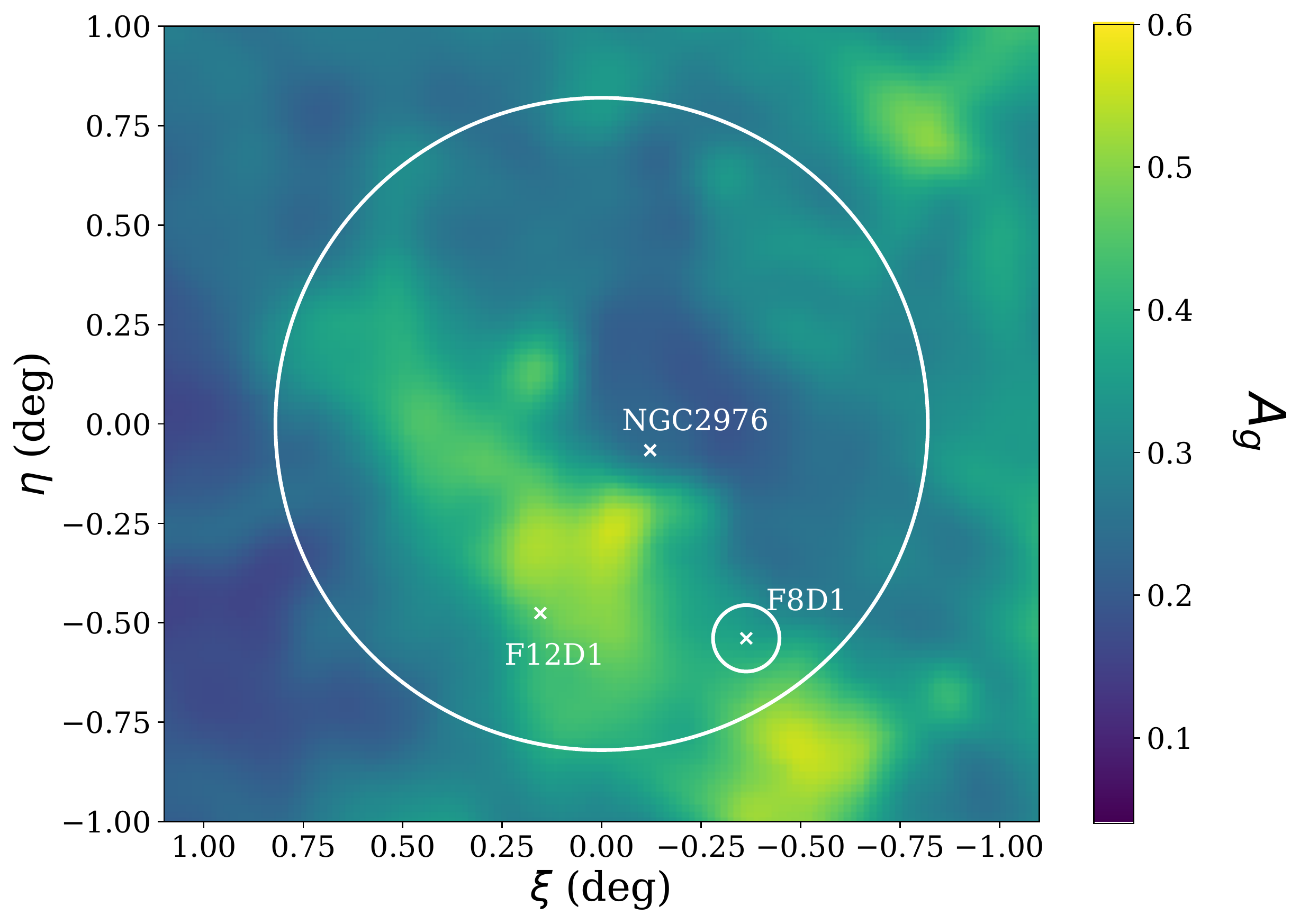}
    \caption{Foreground extinction in the F8D1 HSC field derived from the \citet{Schlegel1998} map, calibrated with the HSC {\it g}-band coefficients from \citet{Schlafly2011}. The large white circle shows the approximate footprint of the HSC pointing analysed here, with the main galaxies marked.  The small white circle delineates 3 $R_{\rm eff}$ around F8D1 derived from our results.}
    \label{fig:extinction}
\end{figure}

The right panel of Fig.\ref{fig:full_map} shows the central high SNR portion of the CFHT stacked $i-$band image (the sky area covered by all 7 exposures from the LSB set), with the main galaxies in the field indicated. This deep image also serves to beautifully illustrate the complex nature of the F8D1 field in terms of galaxy locations and foreground cirrus.  It  demonstrates precisely why the wide-field resolved star approach described in Sec. \ref{hsc} is the only means to study stellar features in the M81 Group at extremely low surface brightness levels.  As expected, the cirrus features in the right panel of Fig.\ref{fig:full_map} generally agree with the areas of high extinction seen in Fig. \ref{fig:extinction}.

\section{A Stream from F8D1}\label{sec:stream}

\begin{figure*}
\includegraphics[width=\textwidth]{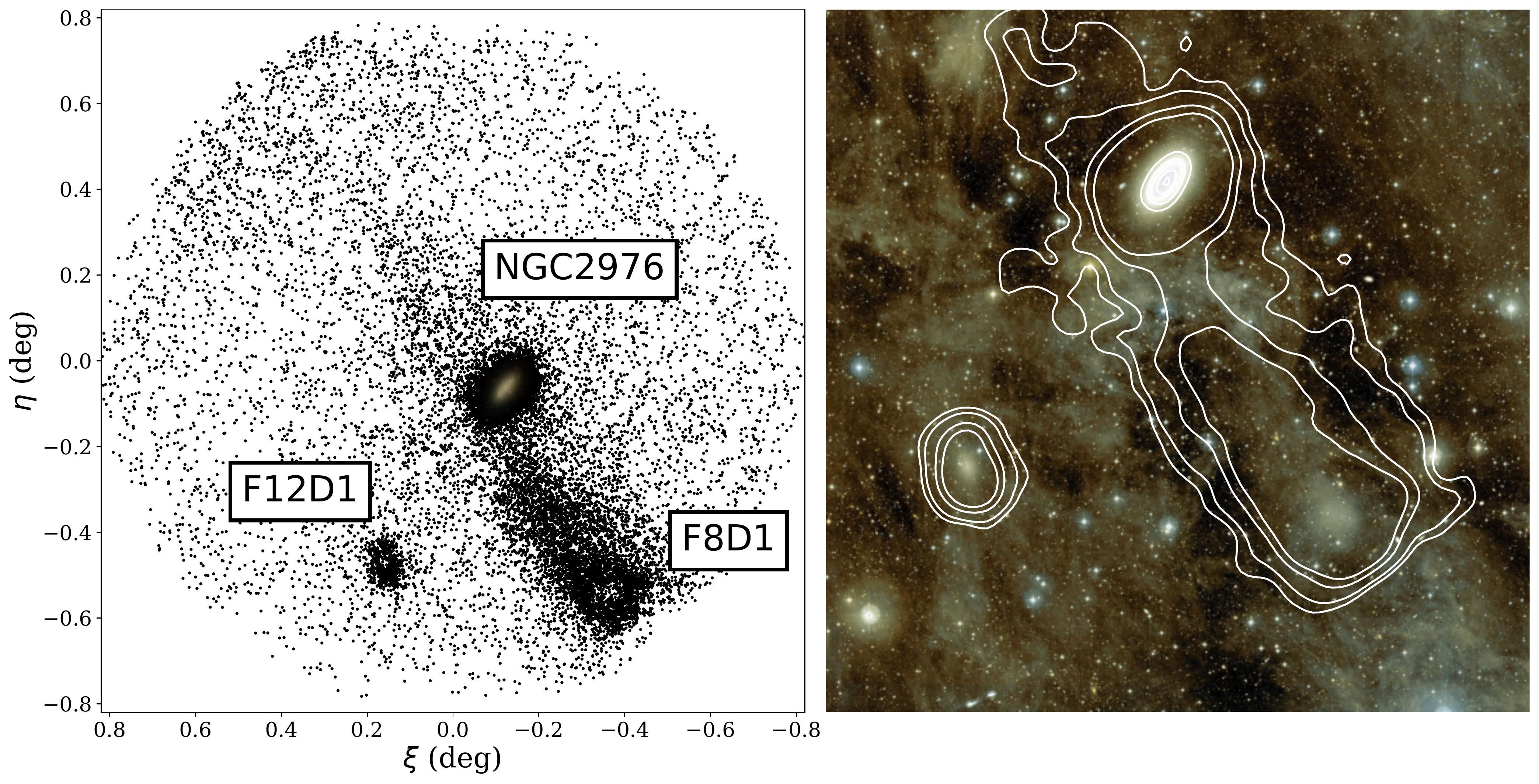}
\caption{Left: The RGB star count density across our HSC pointing.  A giant tidal stream can be seen emanating from F8D1, which is located at the South-Western edge.   The stream  can be traced for more than a degree ($\gtrsim60$~kpc) towards the North-East, in the direction of NGC\,2976 and M81.  A DSS image of NGC\,2976 is superposed for scale and the hollow centres of the galaxies indicate regions where the pipeline failed to return photometry.  Right: A CFHT false-colour image created from our combined $g,r$-band observations with the RGB stellar density contours representing 0.7, 1, 2 and 3 $\sigma$  from the left panel overlaid. Diffuse emission from Galactic cirrus covers the whole field.} \label{fig:starcount}
\end{figure*}

\begin{figure*}
	\includegraphics[width=15cm]{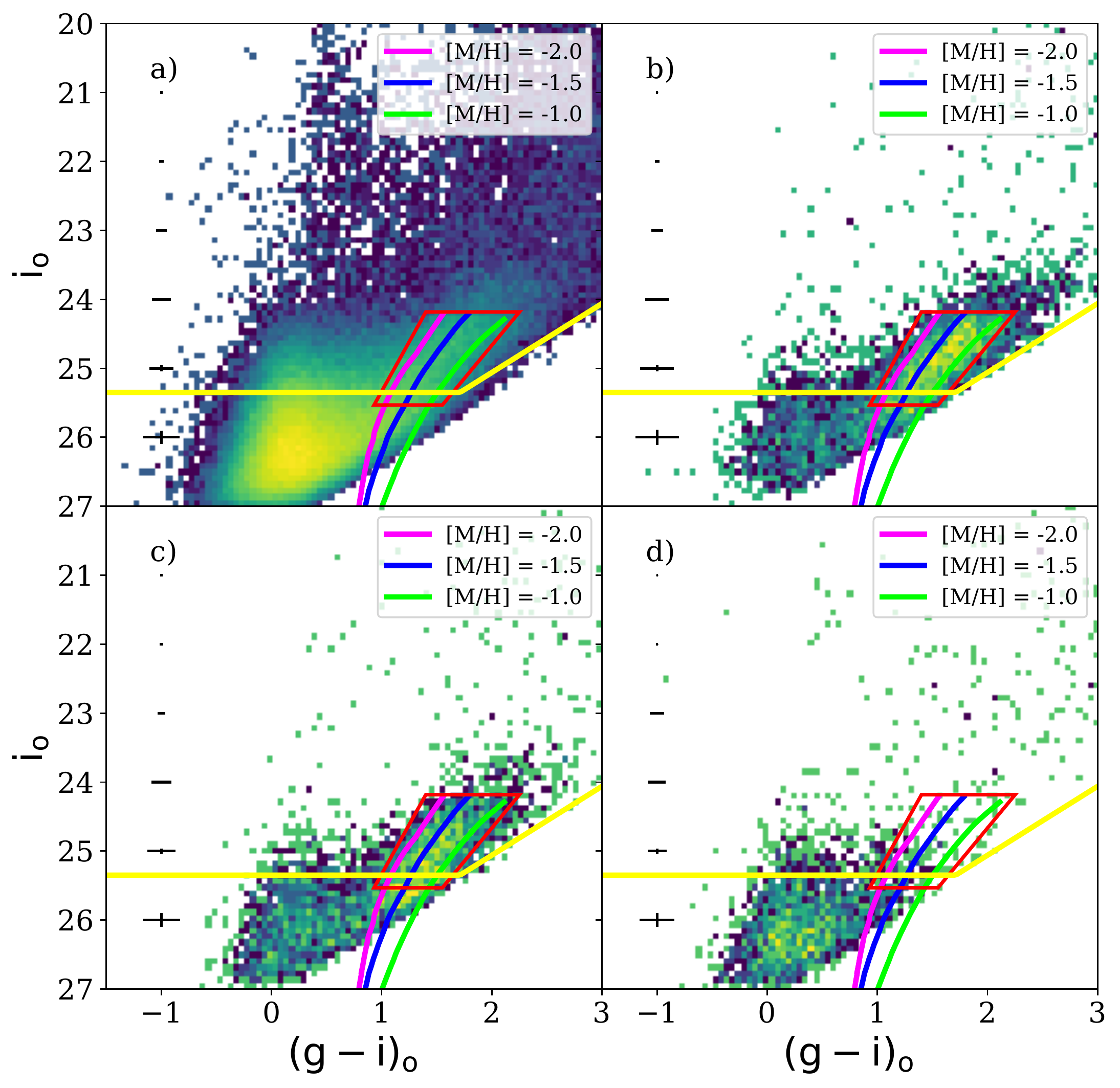}
    \caption{(a): The Hess diagram of stellar sources across the full HSC field. (b):  The Hess diagram of stellar sources within a circular radius of 6 arcmin of F8D1.  (c): The Hess diagram of a region extending from 6 to 24 arcmin away from F8D1, along the direction of the tidal stream. (d): The Hess diagram of a reference field with circular radius 9 arcmin, located well away from the tidal stream and main galaxies. The RGB star selection box (red) and 10 Gyr isochrones of varying metallicity ([M/H] = $-2.0$, $-1.5$, 1.0) are overlaid in each case. Note the RGB populations in the area of F8D1 and its stream are not visible in the reference field. The solid yellow line shows the 50 percent completeness threshold. }
    \label{fig:cmds}
\end{figure*}

Fig. \ref{fig:starcount}  shows the spatial density distribution of candidate RGB stars across the HSC field in a standard coordinate projection anchored to the pointing centre. The sources are selected from our stellar catalogue to lie within a polygonal region on the colour-magnitude diagram (CMD). This region is defined to isolate old metal-poor stars ([M/H]$=-2.0$ to $-1.0$ dex and $\sim$10 Gyr) and extends down to $i = 25.5$, where we have $\gtrsim$50\% completeness (see Fig. \ref{fig:cmds}).  As the three galaxies in our field are all low mass systems, this metallicity range captures the bulk of the RGB stars in the field. 

The most striking feature in this map is a giant stream of stars which extends from F8D1 to the North-West, in the direction of NGC\,2976 and M81.  This feature can be visibly traced for at least a degree on the sky (and $\approx$ two-thirds of the HSC field), and can be seen on both sides of NGC\,2976. Along with F8D1 and its tidal stream, NGC\,2976 and F12D1 also appear as prominent overdensities in this field, both of which have stellar extents that are considerably larger than previously thought \citep{Kara2000, Simon2003}.  To illustrate this, a DSS cutout of the main optical body of NGC\,2976 is superposed on the star count map of Fig. \ref{fig:starcount} (left); this is the only one of the three galaxies that is of high enough surface brightness to be seen in the DSS. All three galaxies have voids in their central regions where the \texttt{hscPipe} failed to return good resolved star photometry due to crowding and the high brightness background. 

The right panel of Fig. \ref{fig:starcount} shows a deep false-colour image created from our combined CFHT $g,r$ images with the contours from the RGB map overlaid. These contours have been created by binning the star count map into 0.5 arcmin bins and gaussian-smoothing with a kernel of 1 arcmin,  and represent levels of 0.7, 1, 2 and 3$\sigma$ above the background.  This visualisation underscores the fact that the tidal tail we have uncovered via star counts is completely impossible to discern on a deep integrated light image due to the bright cirrus in the field. Indeed, while there is an apparent distortion of F8D1 towards the North-East in integrated light (also noted by \citetalias{Caldwell98}), this feature is complicated by Galactic cirrus since its position angle is offset somewhat from that of the tidal stream.  In contrast to F8D1, the outer star count contours of NGC\,2976 and F12D1 are well-behaved and show no evidence that these systems are tidally-distorted in their outer regions.

Fig. \ref{fig:cmds} shows various Hess diagrams constructed from our stellar catalogue with PARSEC isochrones \citep{Bressan2012} with fixed age of 10 Gyr and varying metallicity [M/H] of $-2.0$ to $-1.0$ overlaid. The 50\% completeness level is shown as a yellow line and the polygonal region used to select RGB stars is delineated in red. The top left panel shows sources across the full HSC field. In addition to the conspicuous RGB around 0.7$\lesssim(g-i)_0\lesssim$2.0, the other obvious sequence that can be seen is due to unresolved background galaxies which populate bluer colours, $-0.5\lesssim(g-i)_0\lesssim0.7$. The top right panel shows the Hess diagram for a circular region of radius 6 arcmin which is centered on F8D1, and bottom left panel shows a region extending from 6 to 24 arcmin away from F8D1 along the tidal stream. Compared to the Hess diagram of a similarly-sized reference field (bottom right) that lies far from the main galaxies and tidal stream, these panels show prominent RGBs.  The Hess diagram of F8D1 shows no evidence of recent star formation in the form of young MS stars, although it does reveal a significant population of luminous AGB stars above the RGB tip, as earlier pointed out by \citetalias{Caldwell98}.

\subsection{TRGB Distances}\label{trgb}
While the distances to F8D1 and NGC\,2976 have previously been determined using various HST datasets \citep[e.g.,][]{Caldwell98,Kara2000, Dalcanton2009}, we decided to rederive these in a consistent manner with our HSC dataset.  We use the tip of the RGB (TRGB) method \citep[e.g.,][]{Lee1993, Sakai1996} and follow the procedure described in \citet{Okamoto2019}.  We begin by defining regions around the galaxies that are to be used for analysis.  For F8D1, we select all stars lying within 6.6 arcmin of our adopted center (see Sec. \ref{struc}) and for NGC\,2976 we use stars lying  within a radius of 10 arcmin.  For each system, we select stars with colours in the range 1.4 < $(g-i)_0$ < 2.25 and with photometric errors smaller than 0.5 mag. We then construct their luminosity functions which are convolved with a Sobel filter.  The same method is also applied to stellar sources in empty reference fields of the same area so that we can subtract off the foreground and background contamination.  The highest peaks in the convolved luminosity functions are determined to be  $i_{\rm TRGB}=24.28\pm0.03$ and $24.18\pm0.02$ for F8D1 and NGC\,2976, respectively. 

\citet{Okamoto2019} used PARSEC v1.2 stellar models to explore how the $i$-band absolute magnitude of the TRGB varies with $g-i$ colour assuming metallicities of [M/H] < $-1.0$ and 10-13 Gyr ages.  We measure a median colour of $(g-i)_0 = 1.87\pm0.12$ for TRGB stars in F8D1 which using Eqn. (1) of  \citet{Okamoto2019} translates into $M_{\rm iTRGB}$ = $-3.55\pm0.02$. Therefore, the distance modulus to F8D1 is $(m-M)_0$ = 27.82 $\pm$ 0.03, corresponding to D = 3.67 Mpc $\pm$ 0.06 Mpc.   This is in excellent agreement with the distance moduli of 27.82--28.88  which have been derived from previous HST studies \citep{Kara2000, Ferr2000, Dalcanton2009}. 

Similarly,  we measure a median colour of $(g-i)_0 = 1.89\pm0.08$ for TRGB stars in NGC\,2976 which translates to $M_{\rm iTRGB} = -3.55\pm0.02$ and a distance modulus of NGC\,2976 is $(m-M)_0$ = 27.72 $\pm$ 0.02, or D = 3.50 Mpc $\pm$ 0.04 Mpc. This is also in very good agreement with the literature value of $27.76$ derived by \cite{Dalcanton2009}.  The uncertainties on our measurements include photometric errors, the $M_{\rm iTRGB}$ calibration uncertainty and the reddening uncertainty. We do not include uncertainties arising from the assumption of a single fixed age, or the variance in TRGB luminosity according to different stellar models, which are likely to be far larger.

These measurements confirm that F8D1 and NGC\,2976 lie at different distances, separated along the line-of-sight by $\sim170$ kpc. 
The angular distance between F8D1 and NGC\,2976 is 0.53 degrees, which corresponds to a projected separation of 34 kpc and a 3D distance of 173 kpc.   There are insufficient stars near the TRGB for us to be able to measure a distance gradient along the stream itself but it seems safe to assume that this feature lies well behind NGC\,2976 along the line-of-sight. 

\subsection{Structural Analysis of F8D1}\label{struc}

Before proceeding to analyse the properties of the stream, we revisit the properties of the main body of F8D1 and update its measured structural parameters.  As the stars in the inner parts of the galaxy are poorly detected due to aforementioned issues, a structural analysis using resolved stars is not viable and we need to utilise diffuse light instead.  However, as mentioned in Sec. \ref{cfht},  our implementation of \texttt{hscPipe} does not preserve the LSB emission of galaxies, leaving F8D1 with a large fraction of diffuse light missing in the final stack.  Instead, we investigate the structure of the main body of F8D1 using the MegaCam LSB images. 

\begin{table}
    \begin{threeparttable}[b]
    \setlength{\tabcolsep}{18pt} 
    \centering
    \caption{\label{tab:f8d1_params} Properties of F8D1 measured in this paper.}
   \begin{tabular}{||c c||} 
     \hline\hline
     Parameter & Value \\ 
     \hline
     R.A. (J2000.0) & 09:44:45.95 \\ 
     Dec (J2000.0) & +67:26:27.7 \\
     $M_{g_{0}, {\rm direct}}$ (mag)\tnote{a} & -13.42 $\pm$ 0.03\\
     $M_{g_{0}, {\rm S\acute{e}rsic}}$ (mag)\tnote{b} & -13.70 $\pm$ 0.04\\
     $(g-r)_0$ \tnote{c}& 0.764 $\pm$ 0.003\\
     $(r-i)_0$ \tnote{c}& 0.310 $\pm$ 0.004\\
     $(m-M)_0$ & $27.82 \pm 0.03$ \\
     $D$ (Mpc) & $3.67 \pm 0.06$\\
     $E(B-V)$ & 0.11 \\
     $[$M/H$]$ (dex) & $-1.14 \pm 0.09$\\
     $\sigma[$M/H$]$ (dex) & $0.37$ \\
     PA (deg)\tnote{d} & $90.4\pm 5.1$ \\
     $\epsilon$\tnote{d} &  $0.12\pm 0.01$\\
     $M_*$ ($M_{\odot}$)\tnote{e}& $\sim6.8 \times 10^7 $ \\
     \hline
    \end{tabular}
    \begin{tablenotes}
       \item [a] The measured magnitude of the main body of F8D1 inside a radius of 3.6 arcmin.
       \item [b] The inferred total magnitude of the main body of F8D1 through extrapolation of the Sérsic profile.
       \item [c] The mean colour inside a radius of 3.6 arcmin.
       \item [d] Average over the radial range 0.5$-$1.5 arcmin.
       \item [e] Main body stellar mass calculated using the inferred total magnitude.
     \end{tablenotes}
    \end{threeparttable}
\end{table}

We use the \texttt{AutoProf} software \citep{Stone2021} to construct the surface brightness profile of F8D1.  \texttt{AutoProf} provides an end-to-end pipeline for non-parametric profile extraction, including masking, sky determination, centroiding and isophotal fitting. It can also work in a forced photometry mode, where the parameters fit in one passband can be applied to extract photometry from other passbands.   \texttt{AutoProf} is based on the earlier isophotal-fitting routine of \citet{Jed87} and includes new regularization techniques from machine learning which allow it to deliver more stable and robust profiles. In our analysis, we use it in default mode to fit elliptical isophotes at a series of radii and extract the surface brightness profile. 

Based on visual inspection, the contaminating cirrus is less severe in the $i$-band image (the galaxy gets brighter while the cirrus gets fainter with respect to the higher sky brightness) and hence we decided to use that as our primary image for fitting the isophotes and extracting surface photometry. The ellipse parameters derived from this run of \texttt{AutoProf} were then applied to the $g-$ and $r-$band images.  The centre of F8D1 is chosen to be the point at which the first FFT mode along a circular isophote\footnote{In fact ten circular isophotes are used and averaged, these are chosen to be from $1-10$ times the PSF FWHM so as to evaluate galaxy scale features and not individual peaks.} is minimized. This selects a location at which there is no global direction of increasing brightness. Dwarf galaxies often have clumpy features that are offset from the centre which could distract a conventional peak-finding algorithm, thus the FFT based method is preferred.  The updated central position is provided in Table \ref{tab:f8d1_params} and differs from the \citetalias{Caldwell98} position by $\sim7$ arcsec in RA and $\sim9$ arcsec in Declination. 

Once the centre is selected, \texttt{AutoProf} then fits a series of isophotes which extend until they reach below 2 times the pixel level SNR, which in our case is a semimajor axis of $\sim 1.3'$. When the limit of isophotal fitting is reached, \texttt{AutoProf} can sample the profile further out by assuming the average of the outer fit values of PA and ellipticity (defined as $\epsilon = 1-b/a$) \citep[see][for more details]{Stone2021}. 

To assess the impact of contamination by foreground cirrus in the vicinity of F8D1, the images were compared against WISE 12\,$\mathrm{\mu m}$ observations using the highest resolution maps from \citet{Meisner2014}. We produced cirrus-corrected MegaCam images through an iterative scaling process that flattens the background around F8D1 on the $g,r$ images. The $i$-band was found to be unaffected and so does not require any correction. Comparing the \texttt{AutoProf} photometry extracted from the $g,r$ images with and without this cirrus correction shows no change to the profiles. However, it did reveal a slight artificial boosting of the surface brightness due to cirrus, requiring a correction of +0.22 mag to the $g$-band profile and +0.20 mag to the $r$-band profile. 

\begin{figure}
	\includegraphics[width=\columnwidth]{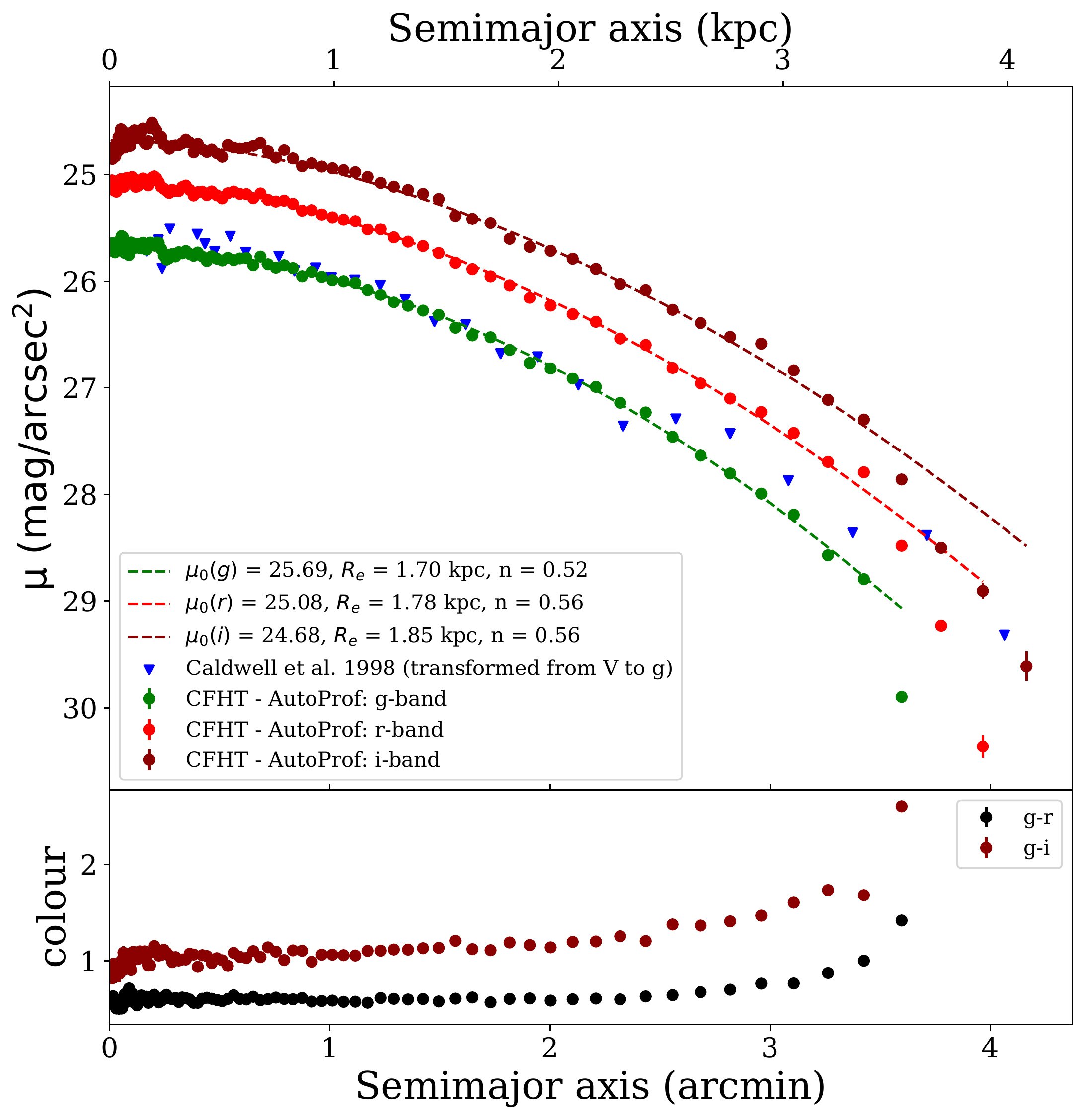}
    \caption{(Top panel) The extinction-corrected, background-subtracted surface brightness profiles of F8D1, derived from a diffuse light analysis of the MegaCam LSB data. Sérsic fits are overlaid. For comparison, we also show the V-band surface brightness profile of  \citetalias{Caldwell98}, which has been converted onto the HSC photometric system and adjusted for our extinction (blue points). (Bottom panel) The extinction-corrected $g-r$ and $g-i$ colour profiles. }
    \label{fig:sersic}
\end{figure}

In Fig. \ref{fig:sersic} we show the three extracted surface brightness profiles (top) and the colour profiles (bottom). These have been corrected for foreground extinction using a single average value of E(B-V)=0.11.  The uncertainties returned by \texttt{Autoprof} are purely statistical in nature. Systematic effects, such as those due to residual cirrus contamination and/or sky residuals, are likely to be several times higher. 
We also overplot the surface brightness of profile of \citetalias{Caldwell98} transformed to the $g-$band using the equations in \citet{Komiyama2018} and adjusted for our assumed extinction. This shows very good agreement with our own profile. The most significant departure is seen at radii $\gtrsim 2$ arcmin, where the \citetalias{Caldwell98}  profile shows a knee in the profile that we do not see in our data.   While there are no strong colour gradients within the inner few arcmin of F8D1, there is a gradual trend to redder colours with increasing radius.  Table \ref{tab:f8d1_params} reports the mean colours of F8D1 measured within 3.6 arcmin. The  ellipticity and position angle derived by  \texttt{Autoprof} are essentially constant over the fit range, averaging to $0.12\pm0.01$ and $90.4\pm5.1$ degrees, respectively. 

\begin{table}
\begin{threeparttable}[b]
    \setlength{\tabcolsep}{1.5pt} 
    \centering
    \caption{\label{tab:f8d1_mainbody} Sérsic fits and derived parameters for the main body of F8D1.}
   \begin{tabular}{||c c c c||} 
     \hline\hline
     Parameter & g & r & i \\ 
     \hline
     $n$ & 0.53 $\pm$ 0.01 & 0.56 $\pm$ 0.01 & 0.56 $\pm$ 0.01 \\
     $\mu(0)$ (mag~arcsec$^{-2}$) & 25.69 $\pm$ 0.01 & 25.08 $\pm$ 0.01  & 24.68 $\pm$ 0.01\\
     $R_{\rm h}$ (arcsec) & 119 $\pm$ 0.6 & 119 $\pm$ 0.8 & 125 $\pm$ 1.1\\
     $R_{\rm eff}$ (arcsec) & 100 $\pm$ 0.8 & 105 $\pm$ 1.2 & 109 $\pm$ 1.6\\
     $R_{\rm eff}$ (kpc) & 1.70 $\pm$ 0.03 & 1.78 $\pm$ 0.04 & 1.85 $\pm$ 0.04\\
     $M_{0,\rm S\acute{e}rsic}$ (mag)\tnote{a} & -13.70 $\pm$ 0.04 & -14.32 $\pm$ 0.04 & -14.82 $\pm$ 0.04\\
     \hline
    \end{tabular}
    \begin{tablenotes}
       \item [a] The inferred total magnitude through extrapolation of the Sérsic profile.
    \end{tablenotes}
    \end{threeparttable}
 \end{table}   
 
We fit a Sérsic profile ($I = I_0 e^{{-(R/R_h)^{\frac{1}{n}}}}$) to the surface brightness profiles, leaving all parameters free \citep{Sersic1968}.  These fits, which are overlaid in Fig. \ref{fig:sersic}, can be seen to provide a good description of the light profile across the main body of F8D1, out to roughly 3.5 arcmin. The values of the central surface brightness, $\mu(0)$, and the semimajor axis scale radius, $R_{\rm h}$,  are reported for each band  in Table \ref{tab:f8d1_mainbody}.  The effective radii, $R_{\rm eff}$,  are also listed, which are obtained from the fit parameters using $R_{\rm eff}= R_{\rm h} b^n$, where $b=2n-0.324$ \citep{Ciotti1991}. Adjusting for our derived distance, we find that the effective radii slightly increase towards redder bands, going from 1.7~kpc in the $g-$band to 1.9~kpc in the $i-$band.  These values lie in between the initial value of 2.45~kpc found by \citetalias{Caldwell98} and the smaller value of 1.2~kpc reported by \citet{Chiboucas2009} from their $r$-band CFHT survey.  When combined our central surface brightnesses of $\mu(0) \sim 24.7-25.7$ mag~arcsec$^{-2}$, it is clear that F8D1 has photometric and structural properties which firmly place it in the realm of UDGs.  

We also revisit the luminosity of F8D1 using two different methods. Firstly, we sum the cumulative flux measured by  \texttt{AutoProf} within the last measured isophote at 3.6 arcmin ($\sim 2.2R_{\rm g,eff}$) and find $M_g=-13.42\pm0.03$.  Alternately, we can use the Sérsic fits to infer the total magnitude using eqn. 12 of \citet{Graham2005}. This yields the slightly larger value of $-13.70$ in the $g$-band (see Table \ref{tab:f8d1_mainbody}),  as expected given that it extrapolates beyond the last directly measured datapoint. We adopt these two measures as the lower and upper limits on the true luminosity of the main body of the system. The latter value compares reasonably  well to the total magnitude reported by \citetalias{Caldwell98} which,  when transformed from $V$ to $g_{\rm hsc}$ and adjusted for our distance and extinction, yields $M_g$ = $-13.9$ mag. With the $i-$band magnitude and $g-i$ colour, we use eqn. 8 of \citet{Taylor2011} to calculate that F8D1 contains $\sim6.8 \times 10^7 M_{\odot}$ in stellar mass.

\subsection{Stream Profile}

\begin{figure}
	\includegraphics[width=\columnwidth]{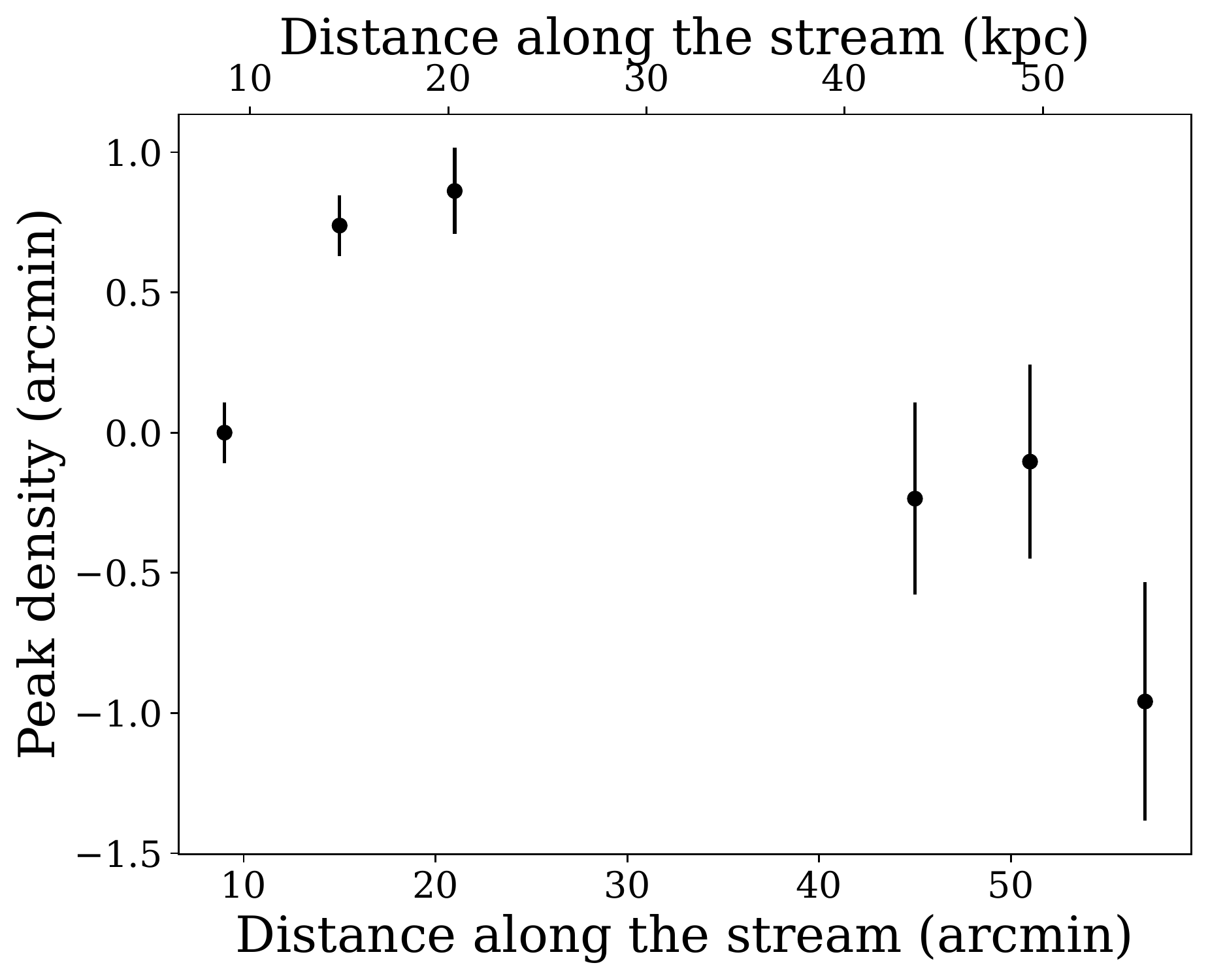}
    \caption{Peak RGB density across the width of the stream, in bins placed sequentially along its length. The locations of the transverse peak densities are measured relative to that of the first bin, which corresponds to $\sim 4R_{\rm g,eff}$ from F8D1. The region corresponding to $\sim$24-42 arcmin along the length of the stream is omitted due to contamination from NGC\,2976. Clear curvature can be seen. }
    \label{fig:curvature}
\end{figure}

Visual inspection of Fig. \ref{fig:full_map} suggests that the tidal tail may have some curvature along its length, as projected on the sky. To quantify this, we consider the RGB star counts in $20\times6$ arcmin wide bins aligned with the direction of the tail. By fitting a Gaussian in the transverse direction, we find the peak RGB star density location in each bin which we take to represent the stream locus (Fig. \ref{fig:curvature}). Bins lying 24 to 42 arcmin along the the stream length are contaminated by stars from NGC\,2976 and are omitted. We find that the stream curves $\mathrm{\sim}$0.8 arcmin ($\sim0.9$~kpc) West of the main body at small radii (6--24 arcmin), and changes direction at larger radii, curving $\mathrm{\sim}$1.1 arcmin ($\sim1.2$~kpc) to the east at distances of 40--60 arcmin.  This is consistent with the S-shaped behaviour that is commonly seen in satellites that are being tidally stripped by a massive companion \citep[e.g.,][]{Johnston2002}.  

\begin{figure}
	\includegraphics[width=\columnwidth]{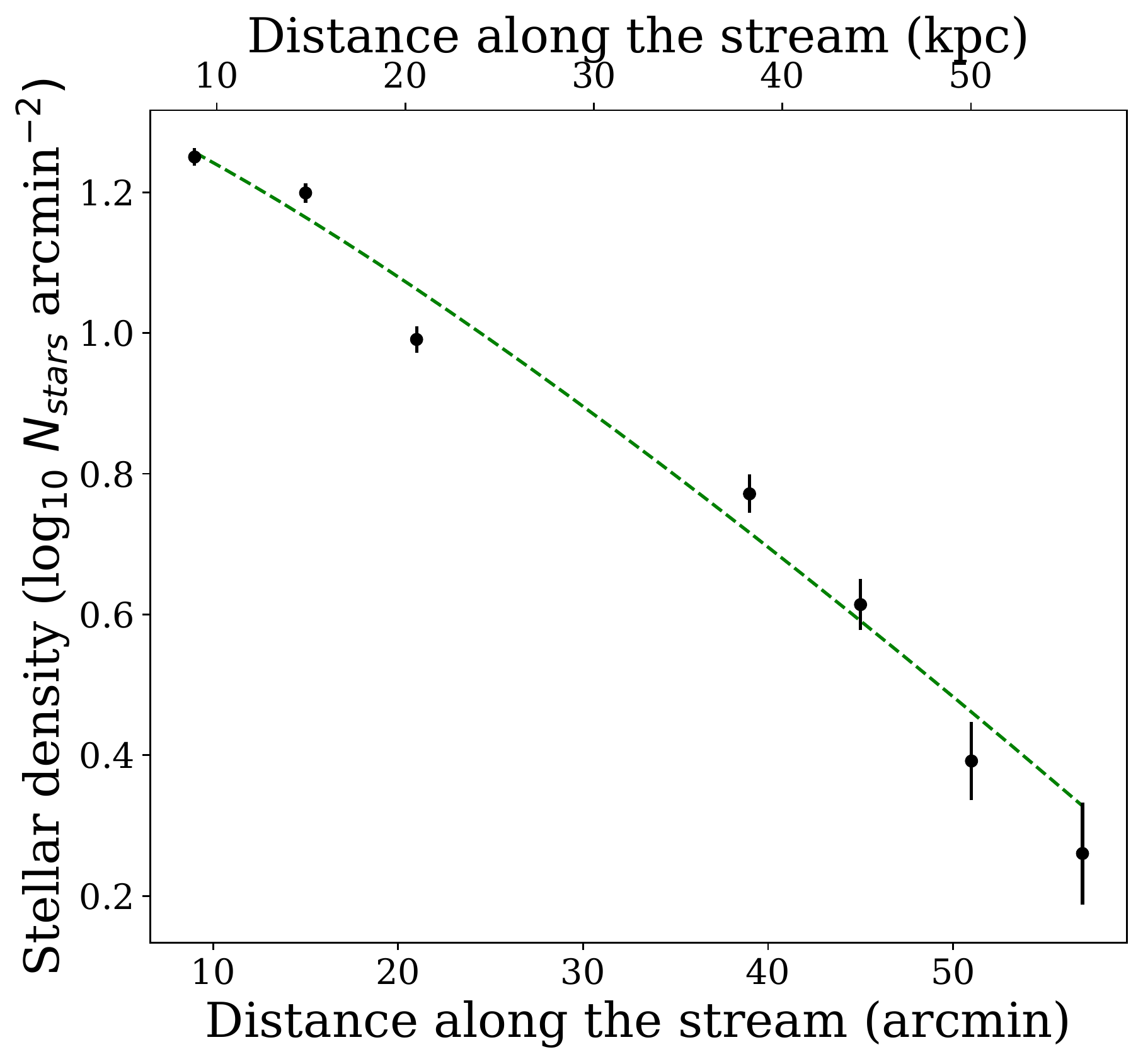}
    \caption{The RGB stellar density profile along the stream length constructed using 8 arcmin wide bins at intervals of 6 arcmin and corrected for contaminants. The bins at $\sim 24-36$ arcmin were omitted due to the presence of NGC\,2976 at this location. The dashed line shows a Sérsic fit to the profile.}
    \label{fig:streamprofile}
\end{figure}

We next investigate the stellar density profile along the tail. For this, we calculate the RGB surface density in $15\times6$ rectangular bins aligned along the direction of the tail. The $\sim15$~kpc width of these bins matches well the visible width of the stream. The bins begin at a distance of 6 arcmin from the centre of F8D1 to avoid the regions where the crowding is severe and completeness corrections are very large.
The distance range 24 -- 36 arcmin is omitted due to the presence of NGC\,2976. The stars counted are completeness corrected. A correction is made for contaminants using the mean source density inside the RGB selection box in nine $12 \times 12$ reference fields positioned far from the obvious overdensities in our HSC pointing. Fig. \ref{fig:streamprofile} shows the resultant profile, with the stream clearly traced to $\approx 60$kpc. The error bars reflect the combination of Poissonian uncertainties on the RGB and contaminant source counts.  The logarithmic stellar density profile can be described reasonably well by a Sérsic model with $R_h \sim 28$ arcmin and  $n \sim 0.8$. Alternately, the outermost four points can be explained by a power-law fall off, with index $\sim -3$. 

We can calculate the luminosity of the stream by summing up the stellar light along its length.  We first consider two large rectangular bins that span 6--24 arcmin and 42--60 arcmin along the tail, each of width 15 arcmin. These bins encompass the visible extent of the stream while avoiding the region contaminated by NGC\,2976. We convert the $g$-band magnitudes of the RGB stars in each bin to flux and sum up the collective light, while also correcting for completeness of each RGB star. We then correct for light from contaminants using the mean contaminant density appropriately scaled to the size of our rectangular bins.   To account for stars with magnitudes fainter than our RGB selection box, we simulate the luminosity function of a 10 Gyr stellar population with [M/H] = $-1.1$ using the Padova isochrones  \citep{Bressan2012}.  Once shifted to the F8D1 distance, the  luminosity function shows that our RGB selection box encompasses 43\% of total luminosity of the population. We therefore divide by this fraction to correct for unresolved stars. Using the average of the luminosities in the two bins as a proxy for that in the 24--42 arcmin bin that is contaminated by NGC\,2976, we calculate that the luminosity contained in the visible stream is $M_g$ = $-12.03$ mag or $7.17\times10^6$ L$_{\odot}$. The luminosity of the main body of F8D1 is estimated to be $2.58-3.34\times10^7$ L$_{\odot}$, depending on whether the directly measured flux inside of 3.6 arcmin or the extrapolated total magnitude is used. These calculations make use of the absolute $g-$band magnitude of the Sun, $M_{g\odot}$ = $5.11$, provided by \citet{Willmer2018}.  Our HSC imagery currently only captures one side of the F8D1 stream but we make the assumption that an analogous tail exists to the South-West. Multiplying the stream luminosity by a factor of two to account for this, we infer that 30-36\% of F8D1's light is contained in the extratidal features. We also calculate that the average surface brightness of the observed stream over the defined area is $\mu_g \sim 32$ mag~arcsec$^{-2}$, which is significantly fainter than any previous integrated light studies of this area have reached. We note that it is not possible to directly join our star count profile of the stream (Fig. \ref{fig:streamprofile}) with our integrated light profile of the main body (Fig. \ref{fig:sersic}) since there is no radial range where the two profiles overlap.

\begin{figure}
	\includegraphics[width=\columnwidth]{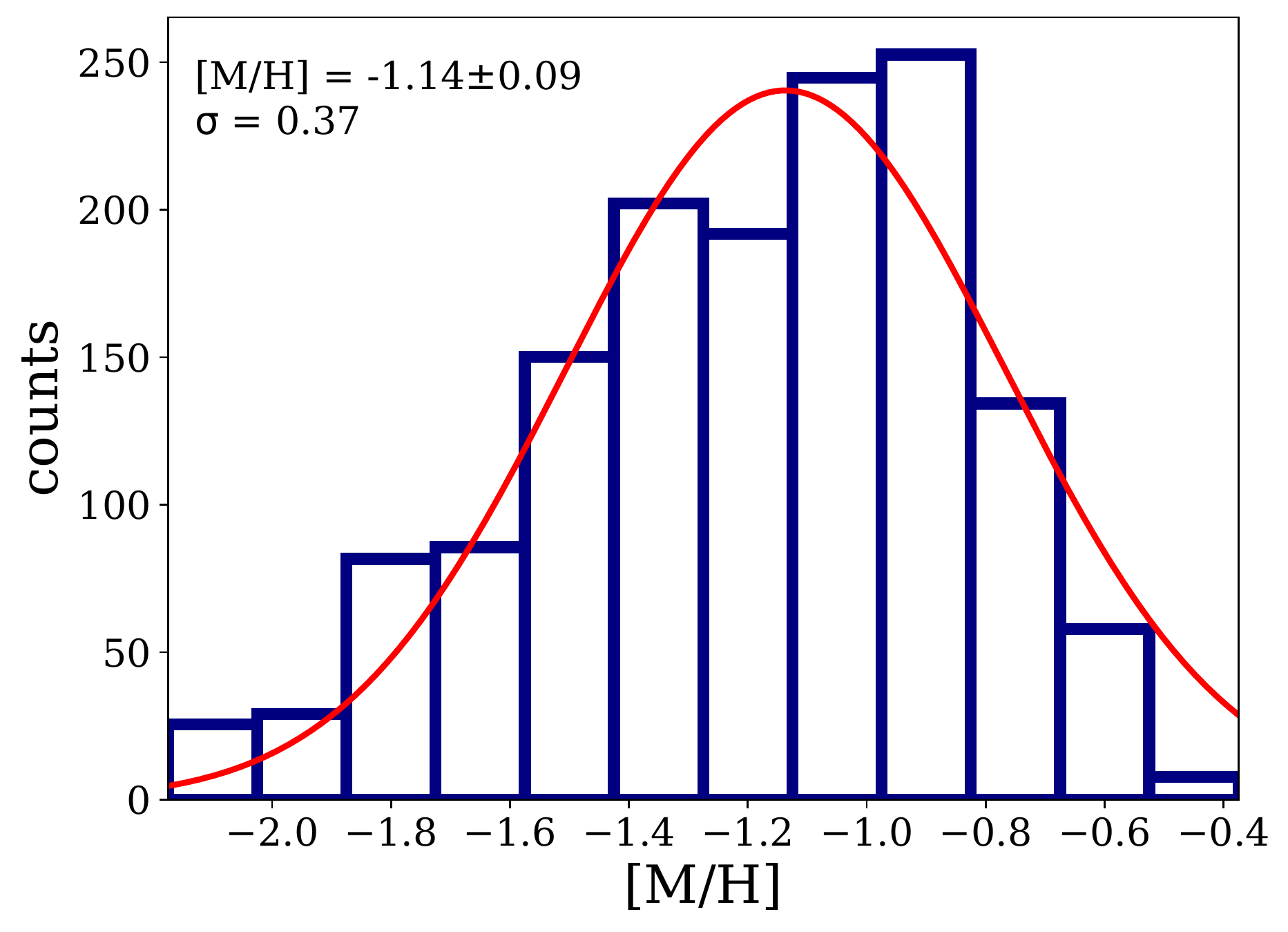}
    \caption{The completeness-corrected and contaminant-subtracted MDF (blue) of the main body of F8D1, constructed using RGB stars lying within $3R_{\rm eff}$ of the galaxy centre.  The MDF for a similar-sized region away from F8D1 is also shown to reflect the contribution of the contaminant population (orange). The peak metallicity is derived from a Gaussian fit to the MDF after subtracting the contribution of contaminants.}
    \label{fig:metallicity}
\end{figure}

\subsection{Metallicity of F8D1 and its stream}
\label{sec:metal}

We investigate the metallicity distribution function (MDF) of resolved stars in the main body and tidal stream of F8D1 using their photometric metallicities ([M/H]). We select PARSEC isochrones of age 10 Gyr and spanning a metallicity range of $-2.2$ to $-0.3$ dex in intervals of 0.05 dex, and shift them to the distance of F8D1.
Each star within our RGB selection box is matched to the closest isochrone in colour (for its magnitude) and assigned that metallicity if the colour difference is $\le 0.05$ mag. The same procedure is performed on a reference field, far removed from the stream,  to allow construction of the MDF of the contaminant population. We note that the metallicities of the contaminant population are meaningless, since these sources are either background galaxies or Milky Way foreground stars and hence not at the distance of F8D1. 

To calculate the main body MDF, we consider RGB stars lying within 
$\sim 3R_{\rm g,eff}$.  Fig. \ref{fig:metallicity} shows the completeness-corrected and contaminant-subtracted MDF (blue)  which exhibits a clear peak and significant dispersion.  The orange  histogram represents the distribution of the contaminants in reference field, scaled to the same area. To characterise the MDF, we fit a Gaussian to estimate the mean metallicity and dispersion which we find to be $\langle[$M/H$]\rangle$ of  $-1.14 \pm0.09$ dex and $\sigma([$M/H$])=0.37$. The errors come from sampling the selected stars from a Gaussian distribution where the width was chosen to be the photometric error of each star centred on the measured magnitude, repeated for N = 2000 times. Our mean metallicity and dispersion compare reasonably well to the values of $\langle[$Fe/H$]\rangle$=$-1.0 \pm0.15$ dex and $\sigma([$Fe/H$])=0.26\pm0.03$ estimated by \citetalias{Caldwell98}, although the metallicity scales are different between these studies.  It should also be kept in mind that our MDF for the main body of F8D1, while based on a sample of several hundred stars, suffers from significant incompleteness. As seen in Fig. \ref{fig:cmds}, this is most likely to affect stars with the reddest colours and hence highest metallicities for a fixed 10 Gyr age. 

\begin{figure}
	\includegraphics[width=\columnwidth]{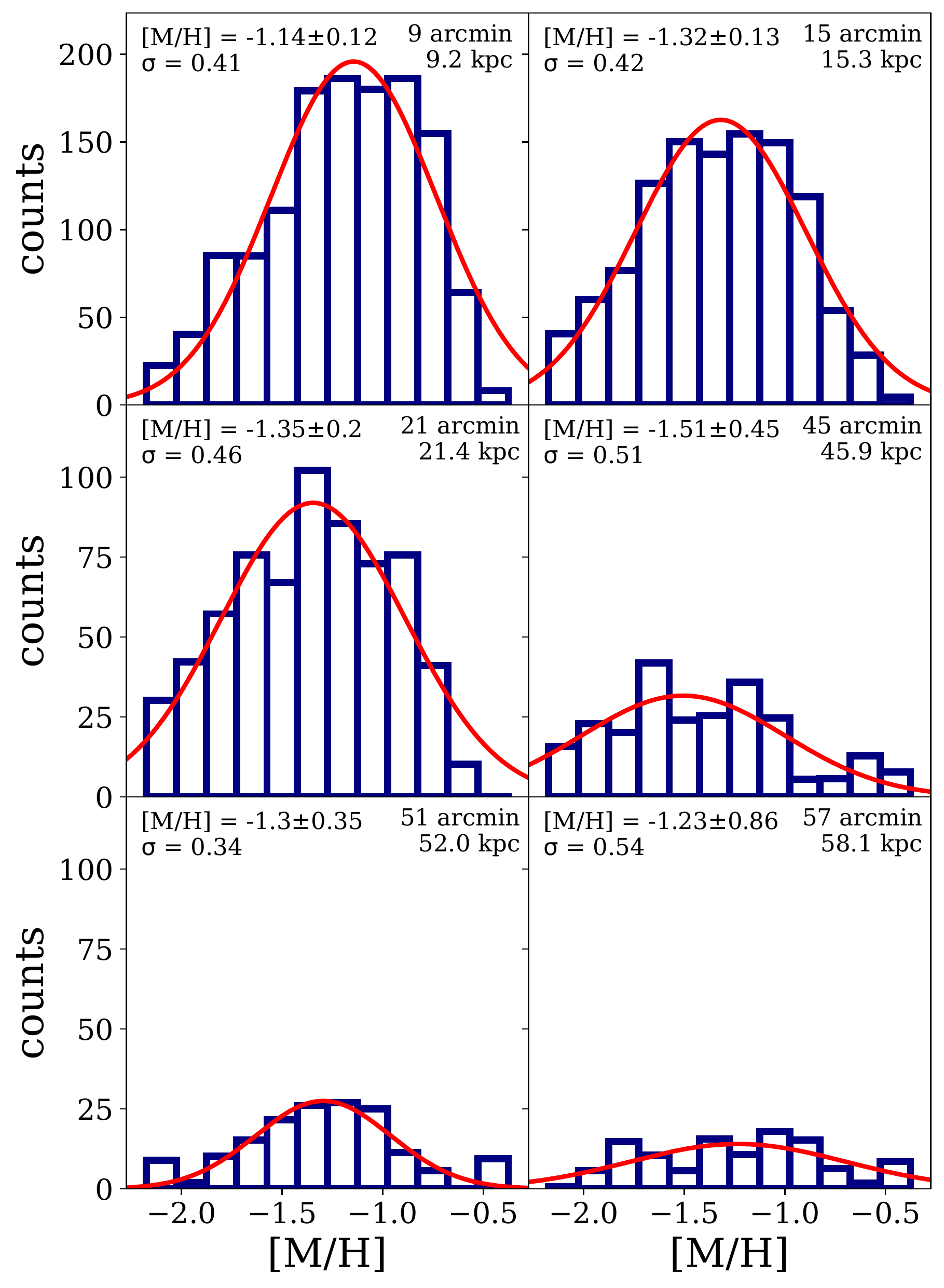}
    \caption{The completeness-corrected and contaminant-subtracted  MDF in bins along the length of the F8D1 stream, constructed in a similar way to that shown in Fig.\ref{fig:metallicity}. In the top left corner of each plot, the mean metallicity is displayed. Below it is the dispersion of the Gaussian fit. On the top right corners is the distance from the main body of F8D1 along the stream in both arcmin and kpc.}
    \label{fig:tail_metal}
\end{figure}

\begin{figure}
	\includegraphics[width=\columnwidth]{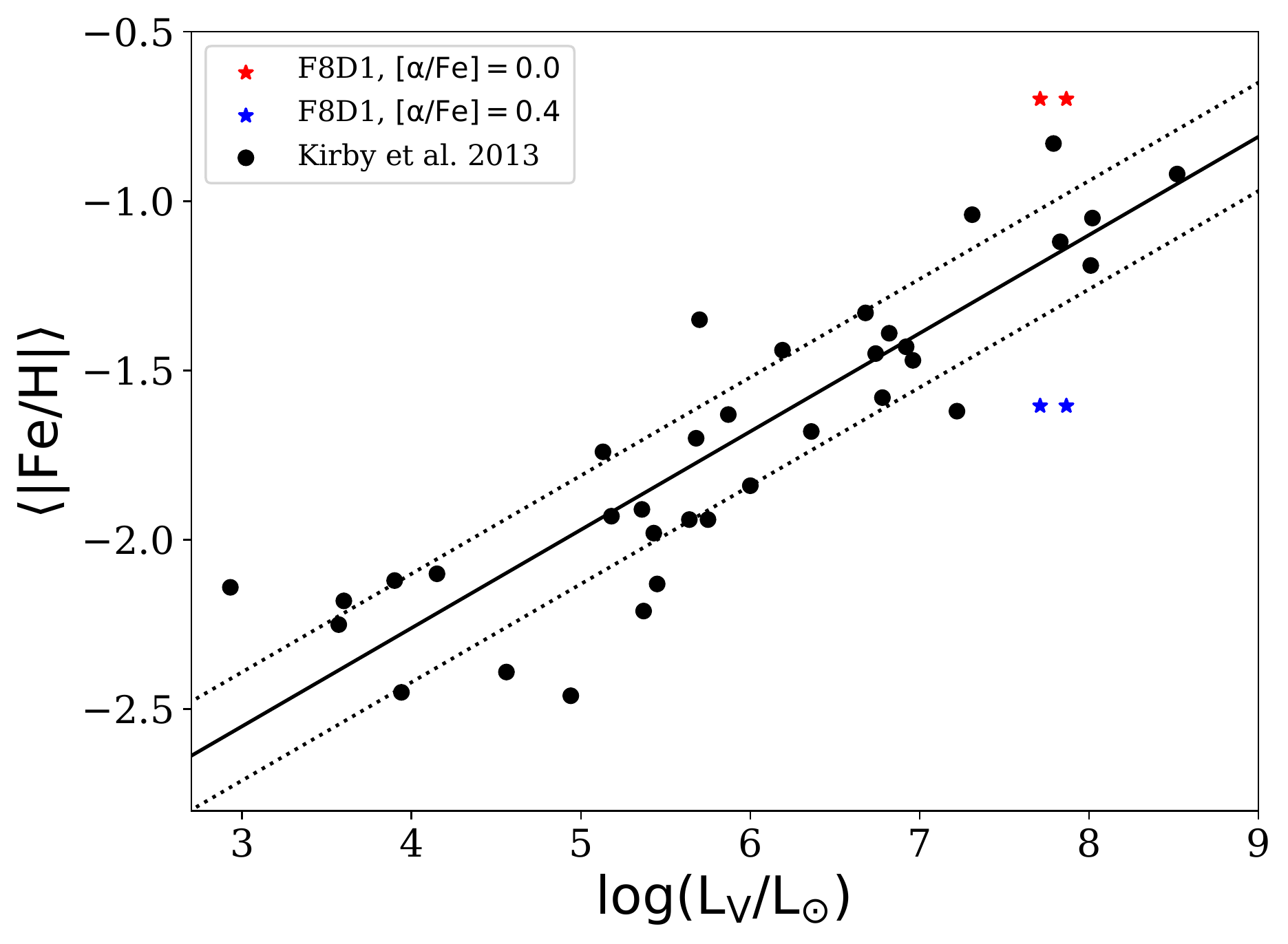}
    \caption{Luminosity-metallicity relation of Local Group dwarf galaxies and F8D1. The black points are taken from \citet{Kirby2013}, including the best-fit and rms lines from eqn. 3 in the paper. The red and blue stars represent F8D1 with $[\alpha/Fe] =$ 0.0 and 0.4 dex, respectively. The lower luminosity points include the main body light only, the higher luminosity represent the main body with added light from the two tidal tails.}
    \label{fig:light_metal}
\end{figure}

Fig. \ref{fig:tail_metal} shows the completeness-corrected and contaminant-subtracted MDFs along the stream, using the same rectangular bins as used to construct the stream radial profile (Sec. \ref{sec:stream}). Again, we use Gaussian fits to characterise the mean metallicity and dispersion in each bin.  Beyond $\sim 20$ kpc, the stream MDFs become increasingly noisy although a clear excess of stars above the expected contaminant population is present in all cases. From inspection of the panels, it is clear that there is no obvious metallicity gradient present. This  suggests that the main body of F8D1 was chemically well-mixed prior to disruption, consistent with the lack of any significant colour gradient in the main body.

Fig. \ref{fig:light_metal} shows where F8D1 lies relative to the luminosity-metallicity relation defined by Local Group dwarfs. We have taken the spectroscopic metallicities and the fit relationship from  \citet{Kirby2013}. The luminosity of F8D1 is transformed from the HSC $g-$band to V using the conversion factors of \citep{Komiyama2018}. To convert our measured [M/H] to [Fe/H], we use eqn. 3 of \citet{Salaris1993} that requires knowledge of the alpha enhancement $\mathrm{[\alpha/Fe]}$. While there are no $\mathrm{[\alpha/Fe]}$ measurements for F8D1, \citet{Ruiz-Lara2018} find that the average $\mathrm{[\alpha/Fe]}$ for their UDGs is $\sim 0.4$. We place F8D1 on this diagram assuming both  $\mathrm{[\alpha/Fe]} \sim 0.0$ (red) and  $\mathrm{[\alpha/Fe]}\sim0.4$ (blue). While F8D1 lies more than  $1 \sigma $ off the relationship in both cases, it falls on different sides of it depending on the assumed $\mathrm{[\alpha/Fe]}$.  This is true whether we consider just the present main body luminosity or the luminosity if we add in the contribution of two tails.  Until a more precise spectroscopic measurement of [Fe/H] can be made for F8D1, the extent to which it remains an outlier on the luminosity-metallicity relation remains inconclusive.

\section{Discussion}
 
We have uncovered a previously-unknown giant tidal stream from the ultra-diffuse galaxy F8D1. Considering its visible extent of $\sim60$ kpc projected on the sky (Fig. \ref{fig:full_map}) and the fact that it contains $\sim$30--36\% of the main body light, it
is clear that the galaxy is undergoing heavy tidal disruption. The two most likely tidal disruptors of F8D1 are its closest projected neighbour, the dwarf spiral NGC\,2976, or the more massive central galaxy M81.  Inspection of HI maps of the M81 Group 
show a large filament pointing in the direction of NGC\,2976 and that the HI distribution of NGC\,2976 is more extended in the direction of M81 \citep{deBlok2018, Sorgho2019}. This suggests a tidal interaction between these two systems but there are no HI features around NGC\,2976 that could be connected to a possible tidal interaction with F8D1. HI has not been detected in F8D1 thus far.  \citet{Sorgho2019} notes that a cloud is visible in the South-East side of NGC\,2976, but it is unlikely to be tied to the disruption of F8D1 located in the South-West.

Our estimated line-of-sight distances to F8D1 and NGC\,2976 show that these galaxies appear near each other only because of their projected positions on the sky.  Their actual 3D separation is 173~kpc, with F8D1 lying further away.  M81 lies at an angular distance of 1.91 degrees from F8D1. Taking this together with a distance of 3.63 Mpc for this system \citep{Jang2012}, we calculate that the 3D distance between M81 and F8D1 is 128 kpc. The total masses of M81 and NGC\,2976 are not well known but their stellar masses have been reported to be $63.8\times10^9$~M$_{\odot}$ and $1.2\times10^9$~M$_{\odot}$ \citep{Sorgho2019}, respectively. The stellar mass-halo mass relation of \citet{Behroozi2013} implies that their halo masses differ by roughly one order of magnitude, and hence that the tidal force exerted on F8D1 by M81 will be at least 20 times stronger than that due to NGC\,2976. The identification of M81 as the primary disruptor is also consistent with the fact that NGC\,2976 shows no obvious distortion in its outermost stellar isophotes.  There is no radial velocity measurement available for F8D1 but given its advanced state of disruption it is natural to surmise that it has had at least one pericentric passage to date. The fact that we see luminous AGB stars in F8D1 (see also \citetalias{Caldwell98}) indicates that star formation has taken place within the last few Gyr. We speculate that  this pericentric passage must have happened fairly recently and that it has been responsible for the stripping the galaxy of its remaining cold gas and quenching star formation. 

There are several interesting and important implications of our findings. Firstly, we can deduce that F8D1 has played a hitherto unrecognised and potentially important role in the interaction history of the M81 Group.  Most famously known for its interacting triplet, all previous studies of the dynamical evolution of the group have focused on the inner trio of M81, M82 and NGC\,3077 alone. These three  galaxies are strongly interacting at the present day, with bridging structures of HI evident between them \citep{Yun1994, deBlok2018} as well as very extended and perturbed stellar halos that are linked by a large stream of tidal debris  \citep{Barker2009, Okamoto2015, Smercina2020}.  

Early modelling of the dynamical evolution of M81, M82 and NGC\,3077 involved studying the pairwise interactions between M81--M82 and M81--NGC\,3077. This suggested pericentric passages about 220 and 280~Myr ago and could match the HI observations reasonably well.  However, this was not full N-body modelling so did not capture the effects of dynamical friction.   More recently, \citet{Oehm2017} have revisited the problem using  N-body modelling. Based on the current positions and velocities of the three inner galaxies, and imposing the requirement that M82 and NGC\,3077 experienced pericentric passages within 30~kpc of M81 in the last 500~Myr, they find it highly unlikely that the current configuration of the system can be long-lived. Indeed, they argue that at least one or both of M82 and NGC\,3077 must have been initially unbound and that we are observing the M81 Group at a special time, with a merger set to take place within the next 1--2~Gyr. On the other hand, for the three galaxies to be in an orbital configuration that is long-lived, they require that the amount of dark matter in the galaxies must be significantly less than would be expected on the basis of their stellar masses. It will be interesting to explore if future modelling efforts that include F8D1 change these conclusions. 

The realisation that F8D1 is in an advanced state of tidal disruption provides an explanation for the peculiar  properties of this system, as first commented on by  \citetalias{Caldwell98}.  In particular, F8D1 has an unusually large size and metallicity for its luminosity and central surface brightness, and \citetalias{Caldwell98} noted that it is  offset from the scaling relationships that are defined by dwarf galaxies in the Local Group.  We have shown that accounting for the visible stream, and assuming that a similar feature lies on the side of the galaxy unprobed by our observations, the luminosity of F8D1 is at least $\sim43-56$\% times larger (corresponding to $M_g \sim -14$ mag) than would be inferred from measuring the main body light. We note that this is likely to be a lower limit on its original luminosity, as it may have experienced repeated earlier bouts of stripping as it has orbited around M81.  This moves the system closer to the expected scaling relation behaviour, suggesting that initially, it may have been a relatively normal galaxy.  

The unveiling of F8D1 as a tidally-disrupted system also raises the question of whether the properties of other UDGs can be explained in this way.  In Table \ref{tab:f8d1_comparison}, we compare the properties of F8D1 that we have derived in this paper with those of two very well-studied UDGs, the systems NGC1052-DF2 and NGC1052-DF4. These systems have attracted much recent attention as examples of UDGs that have little to no dark matter and unusually luminous globular cluster systems  \citep[e.g.,][]{vDokkum2018,vDokkum2019}.  We use the photometric and structural parameters from \citet{Roman2021} for DF2 and DF4, adjusted for their respective distances reported by \citet{Danieli2020} and \citet{Shen2021}. It can be seen that while F8D1 has a somewhat lower central surface brightness and luminosity than these systems, their structural properties compare surprisingly well. In particular, all systems have effective radii $\gtrsim 1.6$~kpc, Sérsic indices smaller than unity and rather circular isophotes in their central regions. Recently \citet{Keim2022} have used new deep imaging data from the Dragonfly Telephoto array to show that NGC1052-DF2 and DF4 exhibit position angle twists and increasingly elongated isophotes when surface brightness levels of $\mu_g \sim 30$ mag~arcsec$^{-2}$ are reached \citep[although see][]{Montes2021}.  While they see no evidence for the strong signatures of ongoing disruption that we have uncovered in F8D1, we note that the F8D1 tail has an even lower mean surface brightness ($\mu_g \sim 32$ mag~arcsec$^{-2}$) than Dragonfly has reached. Indeed,  the F8D1 tail lies well below the surface brightness limit of all integrated light studies of individual UDGs carried out to date\footnote{{\citet{Mowla2017} stack 231 UDGs in the Coma cluster and find no evidence of an S-shaped tidal feature to $\gtrsim 30$ mag~arcsec$^{-2}$. However, this may not be unexpected given that tidal features are likely to exhibit a variety of morphologies and sizes around individual galaxies, which stacking will cancel out.}}.   This leaves open the possibility that many of these systems could have relatively regular shapes and radial profiles in their inner regions, yet still be undergoing significant tidal stripping.  

Tidal stripping and heating have been discussed by several authors as a mechanism for producing UDGs \citep[e.g.,][]{Carleton2019, Jiang2019, Sales2020, Benavides2022}. These papers have mostly focused on galaxies residing in dense environments but they have also discussed UDGs in groups and in the field.  They generally show that  satellite galaxies born in $\sim10^{10}-10^{11}$~M$_{\odot}$ haloes can suffer a dramatic reduction in surface brightness and expansion in size due to tidal stripping and heating.  However, this is not the only way to form UDGs in the simulations, and many are also born as intrinsically low surface brightness extended dwarfs.  \citet{Jiang2019} argue that roughly half of the population of group UDGs have started off as normal dwarfs on highly eccentric orbits. For these systems, one  pericentric passage is sufficient for the system to puff up, and lose its gas due to ram pressure stripping.  The presence of a few Gyr old AGB stars in F8D1 is interesting in this respect, suggesting the pericentric passage around M81 occurred on this timescale. A precise  measurement of the star formation history of F8D1 through modelling its CMD will further constrain this, and will also help distinguish between competing formation scenarios.  For example, \citet{Carleton2019} predict that the mean stellar age for UDGs with effective radii between $1.5-3$~kpc is 4.8~Gyr, while \citet{Sales2020} predict that tidal UDGs should be significantly older.  Thus far, the available age measurements for UDGs are consistent with ages $\gtrsim 7$~Gyr, albeit with large uncertainties \citep[e.g.,][]{Gu2018, FerreMateu2018}. Furthermore, a spectroscopic measurement of F8D1's metallicity will clarify if it really is an outlier on the luminosity-metallicity relationship as would be expected in the tidal-stripping scenario \citep[e.g.,][]{Sales2020}.

\begin{table}
\begin{threeparttable}[b]
    \setlength{\tabcolsep}{3pt} 
    \centering
    \caption{\label{tab:f8d1_comparison} Comparison of F8D1 and NGC1052-DF2 and DF4.}
   \begin{tabular}{||c c c c||} 
     \hline\hline
     Parameter & F8D1 & NGC1052-DF2\tnote{a} & NGC1052-DF4\tnote{b} \\
     \hline
    $R_{\rm eff}$ (kpc) & 1.70 $\pm$ 0.03 & 2.28$\pm$ 0.03 & 1.63$\pm$ 0.04\\
   $\mu_g(0)$ & 25.69 $\pm$ 0.01 & 24.7  $\pm$ 0.1 & 24.3$\pm$ 0.1\\
     $(g-r)_0$ & 0.764 $\pm$ 0.003 & 0.6 $\pm$ 0.01  & 0.63 $\pm$ 0.01\\
     $n$ & 0.52 $\pm$ 0.01 & 0.58$\pm$ 0.02  & 0.8 $\pm$ 0.04 \\
    $b/a$  & 0.88$\pm0.01$  & 0.89 $\pm$ 0.01  & 0.86 $\pm$ 0.01  \\
    M$_{g_{0}}$ (mag) & -13.70 $\pm$ 0.04 & -14.89$\pm$ 0.04  & -14.65$\pm$ 0.04 \\
     \hline
    \end{tabular}
    \begin{tablenotes}
       \item [a] Derived in this paper.
       \item [b] Taken from \citet{Roman2021} and using the \citet{Danieli2020} distance.
       \item [c] Taken from \citet{Roman2021} and using the \citet{Shen2021} distance.
    \end{tablenotes}
    \end{threeparttable}
 \end{table}   

\section{Conclusions}

We have used data from Subaru/HSC and CFHT/MegaCam to revisit the properties of F8D1, a peculiar dwarf satellite companion of M81. Despite being the closest known example of a UDG to the Milky Way, it has been very poorly studied since its discovery more than 20 years ago.  Indeed, the complex and bright Galactic cirrus around F8D1 makes analysis of its faint structure very challenging. 

Using Subaru/HSC observations in the $g$ and $i$ filters, we have mapped the distribution of individual RGB stars on the North-East side of F8D1. We have found evidence for a giant tidal tail emanating from this system. This feature can be traced for over 1 degree on the sky, corresponding to $\gtrsim60$~kpc at the distance of the galaxy. We have mapped the distribution of stars along and across the stream, finding behaviour consistent with an S-shaped structure that is characteristic of tidal stripping. Using the TRGB distance estimation method, we confirm that F8D1 lies on the far side of the M81 Group at D = 3.67 Mpc $\pm$ 0.06 Mpc while the nearby dwarf spiral NGC\,2976 is on the near side at D = 3.50 Mpc $\pm$ 0.04 Mpc. We argue that the most likely cause of F8D1's tidal disruption is the central galaxy M81. 
We calculate that the luminosity contained in the visible stream is $M_g$ = $-12.03$ mag or $7.17\times10^6$ L$_{\odot}$. We determine the photometric metallicity distribution of RGB stars in the main body and stream of F8D1 and find that these are consistent with each other, with no evidence for a significant gradient in either component. The CMD of F8D1 shows no evidence of recent star formation in the form of young MS stars, although it does reveal a significant population of AGB stars. These stars indicate that some star formation took place within the last few Gyr, and their presence may be used to time the last pericentric passage around M81. 
 
 The deep CFHT/MegaCam data allow us to improve the photometric and structural characterisation of the main body of F8D1 and place its tidal stream in context. We find the effective radius to be in the range of 1.7-1.9 kpc, with a trend towards larger values for redder filters.  When combined with the central surface brightnesses of $\mu(0) \sim 24.7-25.7$ mag~arcsec$^{-2}$, it is clear that F8D1 is a {\it bona fide} example of an UDG.  We update the main body luminosity of F8D1 to be $2.58-3.34\times10^7$ L$_{\odot}$, depending on whether the directly measured flux inside of 3.6 arcmin or the extrapolated total magnitude is used. Assuming a  comparable stream on the other side of F8D1 beyond the extent of our current imagery, we deduce that 30-36\% of F8D1's light is contained in the extratidal features. Alternately, accounting for the light in the tails increases the luminosity of F8D1 by 43-56\%, or to $M_g=-14$ mag.   
 
 Revealing F8D1 to be 
 a system in an advanced state of tidal disruption has implications for both the dynamical evolution of the M81 Group and for the origin of galaxies that exhibit UDG properties.  In the first instance, F8D1 has likely played a hitherto unrecognised role in the interaction history of the group and it will be important to include its effects in future modelling efforts. To this end, measuring an accurate radial velocity for this system will be  of great importance. 

 Furthermore,    
 the severe tidal shredding of F8D1 is likely the origin of its present-day extremely diffuse nature, and not any peculiar properties it might have had at birth.   It will be very important to determine the mass-to-light ratio of F8D1 to constrain its dark matter content and compare it to other UDGs. This could be achieved with integrated light spectroscopy of the main body or with the use of dynamical tracers such as GCs, assuming a population of such objects can be found.  
 
 Our results illustrate the critical role that surface brightness limiting depth plays in detecting tidal features around UDGs.  By virtue of F8D1's proximity, we have been able to apply the resolved star approach and thereby detect surface brightness levels of $\mu_g \gtrsim 32$ mag~arcsec$^{-2}$. This is significantly below what has been  achieved in current integrated light studies of other individual UDGs.  As the
only UDG that is close enough to allow studies at extremely low surface brightness, and the first to be unambiguously linked to
tidal stripping, our results for F8D1 are of particular importance.  It leaves open the possibility that many other UDGs could be the result of similar processes, with the most telling signatures of this lurking below current detection limits.
 
 Upcoming large-area sky surveys such as Euclid Wide Survey \citep{Scara2022}, the Legacy Survey of Space and Time with the Vera C. Rubin Observatory \citep{Ivezic2019} and the High Latitude Survey of the Nancy Grace Roman Space Telescope \citep{Akeson2019} will enable giant leaps in the number of systems that can have their large-scale structures and stellar contents analysed via their resolved star populations. These surveys are poised to be transformative for LSB Local Volume science and will yield rich information on the statistical properties of galaxy outskirts within $\sim 5$~Mpc  \citep[e.g.,][]{Pearson2019,Mutlu2021}.  However, as we have demonstrated, the ability to  conduct robust and quantitative studies of LSB diffuse light, either in the high-crowding regions of local galaxies or in galaxies that are too far away for the purely resolved star approach, is also essential. This is a far more challenging requirement and will greatly depend on the adopted strategies for pipeline processing and sky subtraction (A. Watkins, private communication). If excellent performances can be achieved, this will enable unprecedented statistical studies of UDG structures and populations throughout the nearby universe and enable F8D1 to be placed in context. 

\section*{Acknowledgements}

R.Z. acknowledges a studentship received from UKRI STFC and supplementary funding from the Institute for Astronomy at the University of Edinburgh (2145313). We thank Aaron Watkins for  useful discussions on sky subtraction and helpful comments from an anonymous referee. S.O. acknowledges support from JSPS Grant-in-Aid for Scientific Research (18H05875, 20K04031, 20H05855). N.A. thanks the Brain Pool program for financial support, which is funded by the Ministry of Science and ICT through the National Research Foundation of Korea (2018H1D3A2000902). 

The methods described in the paper have been implemented using Python code and several Python libraries: Matplotlib for visualisation \citep{Hunter2007}, NumPy and AstroPy for data handling \citep{Harris2020,Astropy1,Astropy2} and Scipy for model fitting \citep{Scipy2020}. We use \texttt{Autoprof} for the surface brightness analysis \citep{Stone2021}. 

This research is based on observations collected at the Subaru Telescope, which is operated by the National Astronomical Observatory of Japan, and on observations obtained with MegaPrime/MegaCam, a joint project of CFHT and CEA/DAPNIA, at the Canada-France-Hawaii Telescope (CFHT) which is operated by the National Research Council (NRC) of Canada, the Institut National des Science de l'Univers of the Centre National de la Recherche Scientifique (CNRS) of France, and the University of Hawaii.  We are honored and grateful for the opportunity of observing the Universe from Maunakea, which has cultural, historical, and natural significance in Hawaii. 

This paper makes use of software developed for Vera C. Rubin Observatory. We thank the Rubin Observatory for making their code available as free software at http://pipelines.lsst.io/.

The Pan-STARRS1 Surveys (PS1) and the PS1 public science archive have been made possible through contributions by the Institute for Astronomy, the University of Hawaii, the Pan-STARRS Project Office, the Max Planck Society and its participating institutes, the Max Planck Institute for Astronomy, Heidelberg, and the Max Planck Institute for Extraterrestrial Physics, Garching, The Johns Hopkins University, Durham University, the University of Edinburgh, the Queen’s University Belfast, the Harvard-Smithsonian Center for Astrophysics, the Las Cumbres Observatory Global Telescope Network Incorporated, the National Central University of Taiwan, the Space Telescope Science Institute, the National Aeronautics and Space Administration under grant No. NNX08AR22G issued through the Planetary Science Division of the NASA Science Mission Directorate, the National Science Foundation grant No. AST-1238877, the University of Maryland, Eotvos Lorand University (ELTE), the Los Alamos National Laboratory, and the Gordon and Betty Moore Foundation.

For the purpose of open access, the author has applied a Creative Commons Attribution (CC BY) licence to any Author Accepted Manuscript version arising from this submission.

\section*{Data Availability}
The data underlying this article will be shared on reasonable request
to the corresponding author.



\bibliographystyle{mnras}
\bibliography{f8d1}

\begin{thebibliography}{}
\makeatletter
\relax
\def\mn@urlcharsother{\let\do\@makeother \do\$\do\&\do\#\do\^\do\_\do\%\do\~}
\def\mn@doi{\begingroup\mn@urlcharsother \@ifnextchar [ {\mn@doi@}
  {\mn@doi@[]}}
\def\mn@doi@[#1]#2{\def\@tempa{#1}\ifx\@tempa\@empty \href
  {http://dx.doi.org/#2} {doi:#2}\else \href {http://dx.doi.org/#2} {#1}\fi
  \endgroup}
\def\mn@eprint#1#2{\mn@eprint@#1:#2::\@nil}
\def\mn@eprint@arXiv#1{\href {http://arxiv.org/abs/#1} {{\tt arXiv:#1}}}
\def\mn@eprint@dblp#1{\href {http://dblp.uni-trier.de/rec/bibtex/#1.xml}
  {dblp:#1}}
\def\mn@eprint@#1:#2:#3:#4\@nil{\def\@tempa {#1}\def\@tempb {#2}\def\@tempc
  {#3}\ifx \@tempc \@empty \let \@tempc \@tempb \let \@tempb \@tempa \fi \ifx
  \@tempb \@empty \def\@tempb {arXiv}\fi \@ifundefined
  {mn@eprint@\@tempb}{\@tempb:\@tempc}{\expandafter \expandafter \csname
  mn@eprint@\@tempb\endcsname \expandafter{\@tempc}}}

\bibitem[\protect\citeauthoryear{{Akeson} et~al.,}{{Akeson}
  et~al.}{2019}]{Akeson2019}
{Akeson} R.,  et~al., 2019, arXiv e-prints, \href
  {https://ui.adsabs.harvard.edu/abs/2019arXiv190205569A} {p. arXiv:1902.05569}

\bibitem[\protect\citeauthoryear{{Alabi}, {Romanowsky}, {Forbes}, {Brodie}  \&
  {Okabe}}{{Alabi} et~al.}{2020}]{Alabi2020}
{Alabi} A.~B.,  {Romanowsky} A.~J.,  {Forbes} D.~A.,  {Brodie} J.~P.,   {Okabe}
  N.,  2020, \mn@doi [\mnras] {10.1093/mnras/staa1763}, \href
  {https://ui.adsabs.harvard.edu/abs/2020MNRAS.496.3182A} {496, 3182}

\bibitem[\protect\citeauthoryear{{Amorisco} \& {Loeb}}{{Amorisco} \&
  {Loeb}}{2016}]{Amorisco2016}
{Amorisco} N.~C.,  {Loeb} A.,  2016, \mn@doi [\mnras] {10.1093/mnrasl/slw055},
  \href {https://ui.adsabs.harvard.edu/abs/2016MNRAS.459L..51A} {459, L51}

\bibitem[\protect\citeauthoryear{{Amorisco}, {Monachesi}, {Agnello}  \&
  {White}}{{Amorisco} et~al.}{2018}]{Amorisco2018}
{Amorisco} N.~C.,  {Monachesi} A.,  {Agnello} A.,   {White} S.~D.~M.,  2018,
  \mn@doi [\mnras] {10.1093/mnras/sty116}, \href
  {https://ui.adsabs.harvard.edu/abs/2018MNRAS.475.4235A} {475, 4235}

\bibitem[\protect\citeauthoryear{{Astropy Collaboration} et~al.,}{{Astropy
  Collaboration} et~al.}{2013}]{Astropy1}
{Astropy Collaboration} et~al., 2013, \mn@doi [\aap]
  {10.1051/0004-6361/201322068}, \href
  {https://ui.adsabs.harvard.edu/abs/2013A&A...558A..33A} {558, A33}

\bibitem[\protect\citeauthoryear{{Astropy Collaboration} et~al.,}{{Astropy
  Collaboration} et~al.}{2018}]{Astropy2}
{Astropy Collaboration} et~al., 2018, \mn@doi [\aj] {10.3847/1538-3881/aabc4f},
  \href {https://ui.adsabs.harvard.edu/abs/2018AJ....156..123A} {156, 123}

\bibitem[\protect\citeauthoryear{{Axelrod}, {Kantor}, {Lupton}  \&
  {Pierfederici}}{{Axelrod} et~al.}{2010}]{Axelrod2010}
{Axelrod} T.,  {Kantor} J.,  {Lupton} R.~H.,   {Pierfederici} F.,  2010, in
  {Radziwill} N.~M.,  {Bridger} A.,  eds,  Society of Photo-Optical
  Instrumentation Engineers (SPIE) Conference Series Vol. 7740, Software and
  Cyberinfrastructure for Astronomy. p. 774015, \mn@doi{10.1117/12.857297}

\bibitem[\protect\citeauthoryear{{Barker}, {Ferguson}, {Irwin}, {Arimoto}  \&
  {Jablonka}}{{Barker} et~al.}{2009}]{Barker2009}
{Barker} M.~K.,  {Ferguson} A.~M.~N.,  {Irwin} M.,  {Arimoto} N.,   {Jablonka}
  P.,  2009, \mn@doi [\aj] {10.1088/0004-6256/138/5/1469}, \href
  {https://ui.adsabs.harvard.edu/abs/2009AJ....138.1469B} {138, 1469}

\bibitem[\protect\citeauthoryear{{Behroozi}, {Wechsler}  \&
  {Conroy}}{{Behroozi} et~al.}{2013}]{Behroozi2013}
{Behroozi} P.~S.,  {Wechsler} R.~H.,   {Conroy} C.,  2013, \mn@doi [\apj]
  {10.1088/0004-637X/770/1/57}, \href
  {https://ui.adsabs.harvard.edu/abs/2013ApJ...770...57B} {770, 57}

\bibitem[\protect\citeauthoryear{{Benavides}, {Sales}, {Abadi}, {Marinacci},
  {Vogelsberger}  \& {Hernquist}}{{Benavides} et~al.}{2022}]{Benavides2022}
{Benavides} J.~A.,  {Sales} L.~V.,  {Abadi} M.~G.,  {Marinacci} F.,
  {Vogelsberger} M.,   {Hernquist} L.,  2022, arXiv e-prints, \href
  {https://ui.adsabs.harvard.edu/abs/2022arXiv220907539B} {p. arXiv:2209.07539}

\bibitem[\protect\citeauthoryear{{Bennet}, {Sand}, {Zaritsky}, {Crnojevi{\'c}},
  {Spekkens}  \& {Karunakaran}}{{Bennet} et~al.}{2018}]{Bennet2018}
{Bennet} P.,  {Sand} D.~J.,  {Zaritsky} D.,  {Crnojevi{\'c}} D.,  {Spekkens}
  K.,   {Karunakaran} A.,  2018, \mn@doi [\apjl] {10.3847/2041-8213/aadedf},
  \href {https://ui.adsabs.harvard.edu/abs/2018ApJ...866L..11B} {866, L11}

\bibitem[\protect\citeauthoryear{{Bosch} et~al.,}{{Bosch}
  et~al.}{2019}]{Bosch2019}
{Bosch} J.,  et~al., 2019, in {Teuben} P.~J.,  {Pound} M.~W.,  {Thomas} B.~A.,
   {Warner} E.~M.,  eds,  Astronomical Society of the Pacific Conference Series
  Vol. 523, Astronomical Data Analysis Software and Systems XXVII. p.~521
  (\mn@eprint {arXiv} {1812.03248})

\bibitem[\protect\citeauthoryear{{Bressan}, {Marigo}, {Girardi}, {Salasnich},
  {Dal Cero}, {Rubele}  \& {Nanni}}{{Bressan} et~al.}{2012}]{Bressan2012}
{Bressan} A.,  {Marigo} P.,  {Girardi} L.,  {Salasnich} B.,  {Dal Cero} C.,
  {Rubele} S.,   {Nanni} A.,  2012, \mn@doi [\mnras]
  {10.1111/j.1365-2966.2012.21948.x}, \href
  {https://ui.adsabs.harvard.edu/abs/2012MNRAS.427..127B} {427, 127}

\bibitem[\protect\citeauthoryear{{Caldwell} \& {Bothun}}{{Caldwell} \&
  {Bothun}}{1987}]{Caldwell1987}
{Caldwell} N.,  {Bothun} G.~D.,  1987, \mn@doi [\aj] {10.1086/114550}, \href
  {https://ui.adsabs.harvard.edu/abs/1987AJ.....94.1126C} {94, 1126}

\bibitem[\protect\citeauthoryear{{Caldwell}, {Armandroff}, {Da Costa}  \&
  {Seitzer}}{{Caldwell} et~al.}{1998}]{Caldwell98}
{Caldwell} N.,  {Armandroff} T.~E.,  {Da Costa} G.~S.,   {Seitzer} P.,  1998,
  \mn@doi [\aj] {10.1086/300233}, \href
  {https://ui.adsabs.harvard.edu/abs/1998AJ....115..535C} {115, 535}

\bibitem[\protect\citeauthoryear{{Carleton}, {Errani}, {Cooper}, {Kaplinghat},
  {Pe{\~n}arrubia}  \& {Guo}}{{Carleton} et~al.}{2019}]{Carleton2019}
{Carleton} T.,  {Errani} R.,  {Cooper} M.,  {Kaplinghat} M.,  {Pe{\~n}arrubia}
  J.,   {Guo} Y.,  2019, \mn@doi [\mnras] {10.1093/mnras/stz383}, \href
  {https://ui.adsabs.harvard.edu/abs/2019MNRAS.485..382C} {485, 382}

\bibitem[\protect\citeauthoryear{{Chiboucas}, {Karachentsev}  \&
  {Tully}}{{Chiboucas} et~al.}{2009}]{Chiboucas2009}
{Chiboucas} K.,  {Karachentsev} I.~D.,   {Tully} R.~B.,  2009, \mn@doi [\aj]
  {10.1088/0004-6256/137/2/3009}, \href
  {https://ui.adsabs.harvard.edu/abs/2009AJ....137.3009C} {137, 3009}

\bibitem[\protect\citeauthoryear{{Ciotti}}{{Ciotti}}{1991}]{Ciotti1991}
{Ciotti} L.,  1991, \aap, \href
  {https://ui.adsabs.harvard.edu/abs/1991A&A...249...99C} {249, 99}

\bibitem[\protect\citeauthoryear{{Cohen} et~al.,}{{Cohen}
  et~al.}{2018}]{Cohen2018}
{Cohen} Y.,  et~al., 2018, \mn@doi [\apj] {10.3847/1538-4357/aae7c8}, \href
  {https://ui.adsabs.harvard.edu/abs/2018ApJ...868...96C} {868, 96}

\bibitem[\protect\citeauthoryear{{Dalcanton} et~al.,}{{Dalcanton}
  et~al.}{2009}]{Dalcanton2009}
{Dalcanton} J.~J.,  et~al., 2009, \mn@doi [\apjs] {10.1088/0067-0049/183/1/67},
  \href {https://ui.adsabs.harvard.edu/abs/2009ApJS..183...67D} {183, 67}

\bibitem[\protect\citeauthoryear{{Danieli}, {van Dokkum}, {Conroy}, {Abraham}
  \& {Romanowsky}}{{Danieli} et~al.}{2019}]{Danieli2019}
{Danieli} S.,  {van Dokkum} P.,  {Conroy} C.,  {Abraham} R.,   {Romanowsky}
  A.~J.,  2019, \mn@doi [\apjl] {10.3847/2041-8213/ab0e8c}, \href
  {https://ui.adsabs.harvard.edu/abs/2019ApJ...874L..12D} {874, L12}

\bibitem[\protect\citeauthoryear{{Danieli}, {van Dokkum}, {Abraham}, {Conroy},
  {Dolphin}  \& {Romanowsky}}{{Danieli} et~al.}{2020}]{Danieli2020}
{Danieli} S.,  {van Dokkum} P.,  {Abraham} R.,  {Conroy} C.,  {Dolphin} A.~E.,
   {Romanowsky} A.~J.,  2020, \mn@doi [\apjl] {10.3847/2041-8213/ab8dc4}, \href
  {https://ui.adsabs.harvard.edu/abs/2020ApJ...895L...4D} {895, L4}

\bibitem[\protect\citeauthoryear{{Di Cintio}, {Brook}, {Dutton}, {Macci{\`o}},
  {Obreja}  \& {Dekel}}{{Di Cintio} et~al.}{2017}]{diCintio2017}
{Di Cintio} A.,  {Brook} C.~B.,  {Dutton} A.~A.,  {Macci{\`o}} A.~V.,  {Obreja}
  A.,   {Dekel} A.,  2017, \mn@doi [\mnras] {10.1093/mnrasl/slw210}, \href
  {https://ui.adsabs.harvard.edu/abs/2017MNRAS.466L...1D} {466, L1}

\bibitem[\protect\citeauthoryear{{Dolphin}}{{Dolphin}}{2016}]{Dolphin2016}
{Dolphin} A.,  2016, {DOLPHOT: Stellar photometry} (\mn@eprint {ascl}
  {1608.013})

\bibitem[\protect\citeauthoryear{{Doppel}, {Sales}, {Navarro}, {Abadi}, {Peng},
  {Toloba}  \& {Ramos-Almendares}}{{Doppel} et~al.}{2021}]{Doppel2021}
{Doppel} J.~E.,  {Sales} L.~V.,  {Navarro} J.~F.,  {Abadi} M.~G.,  {Peng}
  E.~W.,  {Toloba} E.,   {Ramos-Almendares} F.,  2021, \mn@doi [\mnras]
  {10.1093/mnras/staa3915}, \href
  {https://ui.adsabs.harvard.edu/abs/2021MNRAS.502.1661D} {502, 1661}

\bibitem[\protect\citeauthoryear{{Duc} et~al.,}{{Duc} et~al.}{2015}]{Duc2015}
{Duc} P.-A.,  et~al., 2015, \mn@doi [\mnras] {10.1093/mnras/stu2019}, \href
  {https://ui.adsabs.harvard.edu/abs/2015MNRAS.446..120D} {446, 120}

\bibitem[\protect\citeauthoryear{{Euclid Collaboration} et~al.,}{{Euclid
  Collaboration} et~al.}{2022}]{Scara2022}
{Euclid Collaboration} et~al., 2022, \mn@doi [\aap]
  {10.1051/0004-6361/202141938}, \href
  {https://ui.adsabs.harvard.edu/abs/2022A&A...662A.112E} {662, A112}

\bibitem[\protect\citeauthoryear{{Ferrarese} et~al.,}{{Ferrarese}
  et~al.}{2000}]{Ferr2000}
{Ferrarese} L.,  et~al., 2000, \mn@doi [\apj] {10.1086/308309}, \href
  {https://ui.adsabs.harvard.edu/abs/2000ApJ...529..745F} {529, 745}

\bibitem[\protect\citeauthoryear{{Ferrarese} et~al.,}{{Ferrarese}
  et~al.}{2012}]{Ferrarese2012}
{Ferrarese} L.,  et~al., 2012, \mn@doi [\apjs] {10.1088/0067-0049/200/1/4},
  \href {https://ui.adsabs.harvard.edu/abs/2012ApJS..200....4F} {200, 4}

\bibitem[\protect\citeauthoryear{{Ferr{\'e}-Mateu} et~al.,}{{Ferr{\'e}-Mateu}
  et~al.}{2018}]{FerreMateu2018}
{Ferr{\'e}-Mateu} A.,  et~al., 2018, \mn@doi [\mnras] {10.1093/mnras/sty1597},
  \href {https://ui.adsabs.harvard.edu/abs/2018MNRAS.479.4891F} {479, 4891}

\bibitem[\protect\citeauthoryear{{Forbes}, {Read}, {Gieles}  \&
  {Collins}}{{Forbes} et~al.}{2018}]{Forbes2018}
{Forbes} D.~A.,  {Read} J.~I.,  {Gieles} M.,   {Collins} M. L.~M.,  2018,
  \mn@doi [\mnras] {10.1093/mnras/sty2584}, \href
  {https://ui.adsabs.harvard.edu/abs/2018MNRAS.481.5592F} {481, 5592}

\bibitem[\protect\citeauthoryear{{Forbes}, {Alabi}, {Romanowsky}, {Brodie}  \&
  {Arimoto}}{{Forbes} et~al.}{2020}]{Forbes2020}
{Forbes} D.~A.,  {Alabi} A.,  {Romanowsky} A.~J.,  {Brodie} J.~P.,   {Arimoto}
  N.,  2020, \mn@doi [\mnras] {10.1093/mnras/staa180}, \href
  {https://ui.adsabs.harvard.edu/abs/2020MNRAS.492.4874F} {492, 4874}

\bibitem[\protect\citeauthoryear{{Graham} \& {Driver}}{{Graham} \&
  {Driver}}{2005}]{Graham2005}
{Graham} A.~W.,  {Driver} S.~P.,  2005, \mn@doi [\pasa] {10.1071/AS05001},
  \href {https://ui.adsabs.harvard.edu/abs/2005PASA...22..118G} {22, 118}

\bibitem[\protect\citeauthoryear{{Greco} et~al.,}{{Greco}
  et~al.}{2018}]{Greco2018}
{Greco} J.~P.,  et~al., 2018, \mn@doi [\apj] {10.3847/1538-4357/aab842}, \href
  {https://ui.adsabs.harvard.edu/abs/2018ApJ...857..104G} {857, 104}

\bibitem[\protect\citeauthoryear{{Gu} et~al.,}{{Gu} et~al.}{2018}]{Gu2018}
{Gu} M.,  et~al., 2018, \mn@doi [\apj] {10.3847/1538-4357/aabbae}, \href
  {https://ui.adsabs.harvard.edu/abs/2018ApJ...859...37G} {859, 37}

\bibitem[\protect\citeauthoryear{{Gunn} \& {Stryker}}{{Gunn} \&
  {Stryker}}{1983}]{Gunn1983}
{Gunn} J.~E.,  {Stryker} L.~L.,  1983, \mn@doi [\apjs] {10.1086/190861}, \href
  {https://ui.adsabs.harvard.edu/abs/1983ApJS...52..121G} {52, 121}

\bibitem[\protect\citeauthoryear{{Harris}, {Harris}  \& {Alessi}}{{Harris}
  et~al.}{2013}]{Harris2013}
{Harris} W.~E.,  {Harris} G. L.~H.,   {Alessi} M.,  2013, \mn@doi [\apj]
  {10.1088/0004-637X/772/2/82}, \href
  {https://ui.adsabs.harvard.edu/abs/2013ApJ...772...82H} {772, 82}

\bibitem[\protect\citeauthoryear{Harris et~al.,}{Harris
  et~al.}{2020}]{Harris2020}
Harris C.~R.,  et~al., 2020, \mn@doi [Nature] {10.1038/s41586-020-2649-2}, 585,
  357

\bibitem[\protect\citeauthoryear{Hunter}{Hunter}{2007}]{Hunter2007}
Hunter J.~D.,  2007, \mn@doi [Computing in Science \& Engineering]
  {10.1109/MCSE.2007.55}, 9, 90

\bibitem[\protect\citeauthoryear{{Impey}, {Bothun}  \& {Malin}}{{Impey}
  et~al.}{1988}]{Impey1988}
{Impey} C.,  {Bothun} G.,   {Malin} D.,  1988, \mn@doi [\apj] {10.1086/166500},
  \href {https://ui.adsabs.harvard.edu/abs/1988ApJ...330..634I} {330, 634}

\bibitem[\protect\citeauthoryear{{Iodice} et~al.,}{{Iodice}
  et~al.}{2020}]{Iodice2020}
{Iodice} E.,  et~al., 2020, \mn@doi [\aap] {10.1051/0004-6361/202038523}, \href
  {https://ui.adsabs.harvard.edu/abs/2020A&A...642A..48I} {642, A48}

\bibitem[\protect\citeauthoryear{{Ivezi{\'c}} et~al.,}{{Ivezi{\'c}}
  et~al.}{2019}]{Ivezic2019}
{Ivezi{\'c}} {\v{Z}}.,  et~al., 2019, \mn@doi [\apj]
  {10.3847/1538-4357/ab042c}, \href
  {https://ui.adsabs.harvard.edu/abs/2019ApJ...873..111I} {873, 111}

\bibitem[\protect\citeauthoryear{{Jang}, {Lim}, {Park}  \& {Lee}}{{Jang}
  et~al.}{2012}]{Jang2012}
{Jang} I.~S.,  {Lim} S.,  {Park} H.~S.,   {Lee} M.~G.,  2012, \mn@doi [\apjl]
  {10.1088/2041-8205/751/1/L19}, \href
  {https://ui.adsabs.harvard.edu/abs/2012ApJ...751L..19J} {751, L19}

\bibitem[\protect\citeauthoryear{{Janssens}, {Abraham}, {Brodie}, {Forbes},
  {Romanowsky}  \& {van Dokkum}}{{Janssens} et~al.}{2017}]{Janssens2017}
{Janssens} S.,  {Abraham} R.,  {Brodie} J.,  {Forbes} D.,  {Romanowsky} A.~J.,
   {van Dokkum} P.,  2017, \mn@doi [\apjl] {10.3847/2041-8213/aa667d}, \href
  {https://ui.adsabs.harvard.edu/abs/2017ApJ...839L..17J} {839, L17}

\bibitem[\protect\citeauthoryear{{Jedrzejewski}}{{Jedrzejewski}}{1987}]{Jed87}
{Jedrzejewski} R.~I.,  1987, \mn@doi [\mnras] {10.1093/mnras/226.4.747}, \href
  {https://ui.adsabs.harvard.edu/abs/1987MNRAS.226..747J} {226, 747}

\bibitem[\protect\citeauthoryear{{Jiang}, {Dekel}, {Freundlich}, {Romanowsky},
  {Dutton}, {Macci{\`o}}  \& {Di Cintio}}{{Jiang} et~al.}{2019}]{Jiang2019}
{Jiang} F.,  {Dekel} A.,  {Freundlich} J.,  {Romanowsky} A.~J.,  {Dutton}
  A.~A.,  {Macci{\`o}} A.~V.,   {Di Cintio} A.,  2019, \mn@doi [\mnras]
  {10.1093/mnras/stz1499}, \href
  {https://ui.adsabs.harvard.edu/abs/2019MNRAS.487.5272J} {487, 5272}

\bibitem[\protect\citeauthoryear{{Johnston}, {Choi}  \&
  {Guhathakurta}}{{Johnston} et~al.}{2002}]{Johnston2002}
{Johnston} K.~V.,  {Choi} P.~I.,   {Guhathakurta} P.,  2002, \mn@doi [\aj]
  {10.1086/341040}, \href
  {https://ui.adsabs.harvard.edu/abs/2002AJ....124..127J} {124, 127}

\bibitem[\protect\citeauthoryear{{Junais} et~al.,}{{Junais}
  et~al.}{2021}]{Junais2021}
{Junais} et~al., 2021, \mn@doi [\aap] {10.1051/0004-6361/202040185}, \href
  {https://ui.adsabs.harvard.edu/abs/2021A&A...650A..99J} {650, A99}

\bibitem[\protect\citeauthoryear{{Juri{\'c}} et~al.,}{{Juri{\'c}}
  et~al.}{2017}]{Juric2017}
{Juri{\'c}} M.,  et~al., 2017, in {Lorente} N.~P.~F.,  {Shortridge} K.,
  {Wayth} R.,  eds,  Astronomical Society of the Pacific Conference Series Vol.
  512, Astronomical Data Analysis Software and Systems XXV. p.~279 (\mn@eprint
  {arXiv} {1512.07914})

\bibitem[\protect\citeauthoryear{{Karachentsev} et~al.,}{{Karachentsev}
  et~al.}{2000}]{Kara2000}
{Karachentsev} I.~D.,  et~al., 2000, \aap, \href
  {https://ui.adsabs.harvard.edu/abs/2000A&A...363..117K} {363, 117}

\bibitem[\protect\citeauthoryear{{Keim} et~al.,}{{Keim}
  et~al.}{2022}]{Keim2022}
{Keim} M.~A.,  et~al., 2022, \mn@doi [\apj] {10.3847/1538-4357/ac7dab}, \href
  {https://ui.adsabs.harvard.edu/abs/2022ApJ...935..160K} {935, 160}

\bibitem[\protect\citeauthoryear{{Kirby}, {Cohen}, {Guhathakurta}, {Cheng},
  {Bullock}  \& {Gallazzi}}{{Kirby} et~al.}{2013}]{Kirby2013}
{Kirby} E.~N.,  {Cohen} J.~G.,  {Guhathakurta} P.,  {Cheng} L.,  {Bullock}
  J.~S.,   {Gallazzi} A.,  2013, \mn@doi [\apj] {10.1088/0004-637X/779/2/102},
  \href {https://ui.adsabs.harvard.edu/abs/2013ApJ...779..102K} {779, 102}

\bibitem[\protect\citeauthoryear{{Koda}, {Yagi}, {Yamanoi}  \&
  {Komiyama}}{{Koda} et~al.}{2015}]{Koda2015}
{Koda} J.,  {Yagi} M.,  {Yamanoi} H.,   {Komiyama} Y.,  2015, \mn@doi [\apjl]
  {10.1088/2041-8205/807/1/L2}, \href
  {https://ui.adsabs.harvard.edu/abs/2015ApJ...807L...2K} {807, L2}

\bibitem[\protect\citeauthoryear{{Komiyama} et~al.,}{{Komiyama}
  et~al.}{2018}]{Komiyama2018}
{Komiyama} Y.,  et~al., 2018, \mn@doi [\apj] {10.3847/1538-4357/aaa129}, \href
  {https://ui.adsabs.harvard.edu/abs/2018ApJ...853...29K} {853, 29}

\bibitem[\protect\citeauthoryear{{Lee}, {Freedman}  \& {Madore}}{{Lee}
  et~al.}{1993}]{Lee1993}
{Lee} M.~G.,  {Freedman} W.~L.,   {Madore} B.~F.,  1993, \mn@doi [\apj]
  {10.1086/173334}, \href
  {https://ui.adsabs.harvard.edu/abs/1993ApJ...417..553L} {417, 553}

\bibitem[\protect\citeauthoryear{{Leisman} et~al.,}{{Leisman}
  et~al.}{2017}]{Leisman2017}
{Leisman} L.,  et~al., 2017, \mn@doi [\apj] {10.3847/1538-4357/aa7575}, \href
  {https://ui.adsabs.harvard.edu/abs/2017ApJ...842..133L} {842, 133}

\bibitem[\protect\citeauthoryear{{Lim}, {Peng}, {C{\^o}t{\'e}}, {Sales}, {den
  Brok}, {Blakeslee}  \& {Guhathakurta}}{{Lim} et~al.}{2018}]{Lim2018}
{Lim} S.,  {Peng} E.~W.,  {C{\^o}t{\'e}} P.,  {Sales} L.~V.,  {den Brok} M.,
  {Blakeslee} J.~P.,   {Guhathakurta} P.,  2018, \mn@doi [\apj]
  {10.3847/1538-4357/aacb81}, \href
  {https://ui.adsabs.harvard.edu/abs/2018ApJ...862...82L} {862, 82}

\bibitem[\protect\citeauthoryear{{Magnier} et~al.,}{{Magnier}
  et~al.}{2013}]{Magnier2013}
{Magnier} E.~A.,  et~al., 2013, \mn@doi [\apjs] {10.1088/0067-0049/205/2/20},
  \href {https://ui.adsabs.harvard.edu/abs/2013ApJS..205...20M} {205, 20}

\bibitem[\protect\citeauthoryear{{Martin} et~al.,}{{Martin}
  et~al.}{2016}]{Martin2016}
{Martin} N.~F.,  et~al., 2016, \mn@doi [\apj] {10.3847/1538-4357/833/2/167},
  \href {https://ui.adsabs.harvard.edu/abs/2016ApJ...833..167M} {833, 167}

\bibitem[\protect\citeauthoryear{{Meisner} \& {Finkbeiner}}{{Meisner} \&
  {Finkbeiner}}{2014}]{Meisner2014}
{Meisner} A.~M.,  {Finkbeiner} D.~P.,  2014, \mn@doi [\apj]
  {10.1088/0004-637X/781/1/5}, \href
  {https://ui.adsabs.harvard.edu/abs/2014ApJ...781....5M} {781, 5}

\bibitem[\protect\citeauthoryear{{Merritt}, {van Dokkum}, {Danieli}, {Abraham},
  {Zhang}, {Karachentsev}  \& {Makarova}}{{Merritt} et~al.}{2016}]{Merritt2016}
{Merritt} A.,  {van Dokkum} P.,  {Danieli} S.,  {Abraham} R.,  {Zhang} J.,
  {Karachentsev} I.~D.,   {Makarova} L.~N.,  2016, \mn@doi [\apj]
  {10.3847/1538-4357/833/2/168}, \href
  {https://ui.adsabs.harvard.edu/abs/2016ApJ...833..168M} {833, 168}

\bibitem[\protect\citeauthoryear{{Mihos} et~al.,}{{Mihos}
  et~al.}{2015}]{Mihos2015}
{Mihos} J.~C.,  et~al., 2015, \mn@doi [\apjl] {10.1088/2041-8205/809/2/L21},
  \href {https://ui.adsabs.harvard.edu/abs/2015ApJ...809L..21M} {809, L21}

\bibitem[\protect\citeauthoryear{{Miyazaki} et~al.,}{{Miyazaki}
  et~al.}{2018}]{Miyazaki2018}
{Miyazaki} S.,  et~al., 2018, \mn@doi [\pasj] {10.1093/pasj/psx063}, \href
  {https://ui.adsabs.harvard.edu/abs/2018PASJ...70S...1M} {70, S1}

\bibitem[\protect\citeauthoryear{{Montes}, {Trujillo}, {Infante-Sainz},
  {Monelli}  \& {Borlaff}}{{Montes} et~al.}{2021}]{Montes2021}
{Montes} M.,  {Trujillo} I.,  {Infante-Sainz} R.,  {Monelli} M.,   {Borlaff}
  A.~S.,  2021, \mn@doi [\apj] {10.3847/1538-4357/ac0d55}, \href
  {https://ui.adsabs.harvard.edu/abs/2021ApJ...919...56M} {919, 56}

\bibitem[\protect\citeauthoryear{{Mowla}, {van Dokkum}, {Merritt}, {Abraham},
  {Yagi}  \& {Koda}}{{Mowla} et~al.}{2017}]{Mowla2017}
{Mowla} L.,  {van Dokkum} P.,  {Merritt} A.,  {Abraham} R.,  {Yagi} M.,
  {Koda} J.,  2017, \mn@doi [\apj] {10.3847/1538-4357/aa961b}, \href
  {https://ui.adsabs.harvard.edu/abs/2017ApJ...851...27M} {851, 27}

\bibitem[\protect\citeauthoryear{{M{\"u}ller}, {Jerjen}  \&
  {Binggeli}}{{M{\"u}ller} et~al.}{2018}]{Muller2018}
{M{\"u}ller} O.,  {Jerjen} H.,   {Binggeli} B.,  2018, \mn@doi [\aap]
  {10.1051/0004-6361/201832897}, \href
  {https://ui.adsabs.harvard.edu/abs/2018A&A...615A.105M} {615, A105}

\bibitem[\protect\citeauthoryear{{Mutlu-Pakdil} et~al.,}{{Mutlu-Pakdil}
  et~al.}{2021}]{Mutlu2021}
{Mutlu-Pakdil} B.,  et~al., 2021, \mn@doi [\apj] {10.3847/1538-4357/ac0db8},
  \href {https://ui.adsabs.harvard.edu/abs/2021ApJ...918...88M} {918, 88}

\bibitem[\protect\citeauthoryear{{Oehm}, {Thies}  \& {Kroupa}}{{Oehm}
  et~al.}{2017}]{Oehm2017}
{Oehm} W.,  {Thies} I.,   {Kroupa} P.,  2017, \mn@doi [\mnras]
  {10.1093/mnras/stw3381}, \href
  {https://ui.adsabs.harvard.edu/abs/2017MNRAS.467..273O} {467, 273}

\bibitem[\protect\citeauthoryear{{Okamoto}, {Arimoto}, {Ferguson}, {Bernard},
  {Irwin}, {Yamada}  \& {Utsumi}}{{Okamoto} et~al.}{2015}]{Okamoto2015}
{Okamoto} S.,  {Arimoto} N.,  {Ferguson} A. M.~N.,  {Bernard} E.~J.,  {Irwin}
  M.~J.,  {Yamada} Y.,   {Utsumi} Y.,  2015, \mn@doi [\apjl]
  {10.1088/2041-8205/809/1/L1}, \href
  {https://ui.adsabs.harvard.edu/abs/2015ApJ...809L...1O} {809, L1}

\bibitem[\protect\citeauthoryear{{Okamoto}, {Arimoto}, {Ferguson}, {Irwin},
  {Bernard}  \& {Utsumi}}{{Okamoto} et~al.}{2019}]{Okamoto2019}
{Okamoto} S.,  {Arimoto} N.,  {Ferguson} A. M.~N.,  {Irwin} M.~J.,  {Bernard}
  E.~J.,   {Utsumi} Y.,  2019, \mn@doi [\apj] {10.3847/1538-4357/ab44a7}, \href
  {https://ui.adsabs.harvard.edu/abs/2019ApJ...884..128O} {884, 128}

\bibitem[\protect\citeauthoryear{{Pearson}, {Starkenburg}, {Johnston},
  {Williams}, {Ibata}  \& {Khan}}{{Pearson} et~al.}{2019}]{Pearson2019}
{Pearson} S.,  {Starkenburg} T.~K.,  {Johnston} K.~V.,  {Williams} B.~F.,
  {Ibata} R.~A.,   {Khan} R.,  2019, \mn@doi [\apj] {10.3847/1538-4357/ab3e06},
  \href {https://ui.adsabs.harvard.edu/abs/2019ApJ...883...87P} {883, 87}

\bibitem[\protect\citeauthoryear{{Pickles}}{{Pickles}}{1998}]{Pickles1998}
{Pickles} A.~J.,  1998, \mn@doi [\pasp] {10.1086/316197}, \href
  {https://ui.adsabs.harvard.edu/abs/1998PASP..110..863P} {110, 863}

\bibitem[\protect\citeauthoryear{{Prole}, {van der Burg}, {Hilker}  \&
  {Davies}}{{Prole} et~al.}{2019}]{Prole2019}
{Prole} D.~J.,  {van der Burg} R.~F.~J.,  {Hilker} M.,   {Davies} J.~I.,  2019,
  \mn@doi [\mnras] {10.1093/mnras/stz1843}, \href
  {https://ui.adsabs.harvard.edu/abs/2019MNRAS.488.2143P} {488, 2143}

\bibitem[\protect\citeauthoryear{{Pucha} et~al.,}{{Pucha}
  et~al.}{2019}]{Pucha2019}
{Pucha} R.,  et~al., 2019, \mn@doi [\apj] {10.3847/1538-4357/ab29fb}, \href
  {https://ui.adsabs.harvard.edu/abs/2019ApJ...880..104P} {880, 104}

\bibitem[\protect\citeauthoryear{{Radburn-Smith} et~al.,}{{Radburn-Smith}
  et~al.}{2011}]{Radburn-Smith2011}
{Radburn-Smith} D.~J.,  et~al., 2011, \mn@doi [\apjs]
  {10.1088/0067-0049/195/2/18}, \href
  {https://ui.adsabs.harvard.edu/abs/2011ApJS..195...18R} {195, 18}

\bibitem[\protect\citeauthoryear{{Regnault} et~al.,}{{Regnault}
  et~al.}{2009}]{Regnault2009}
{Regnault} N.,  et~al., 2009, \mn@doi [\aap] {10.1051/0004-6361/200912446},
  \href {https://ui.adsabs.harvard.edu/abs/2009A&A...506..999R} {506, 999}

\bibitem[\protect\citeauthoryear{{Rodrigo} \& {Solano}}{{Rodrigo} \&
  {Solano}}{2020}]{Rodrigo2020}
{Rodrigo} C.,  {Solano} E.,  2020, in XIV.0 Scientific Meeting (virtual) of the
  Spanish Astronomical Society. p.~182

\bibitem[\protect\citeauthoryear{{Rom{\'a}n} \& {Trujillo}}{{Rom{\'a}n} \&
  {Trujillo}}{2017}]{Roman2017}
{Rom{\'a}n} J.,  {Trujillo} I.,  2017, \mn@doi [\mnras] {10.1093/mnras/stx694},
  \href {https://ui.adsabs.harvard.edu/abs/2017MNRAS.468.4039R} {468, 4039}

\bibitem[\protect\citeauthoryear{{Rom{\'a}n}, {Castilla}  \&
  {Pascual-Granado}}{{Rom{\'a}n} et~al.}{2021}]{Roman2021}
{Rom{\'a}n} J.,  {Castilla} A.,   {Pascual-Granado} J.,  2021, \mn@doi [\aap]
  {10.1051/0004-6361/202142161}, \href
  {https://ui.adsabs.harvard.edu/abs/2021A&A...656A..44R} {656, A44}

\bibitem[\protect\citeauthoryear{{Ruiz-Lara} et~al.,}{{Ruiz-Lara}
  et~al.}{2018}]{Ruiz-Lara2018}
{Ruiz-Lara} T.,  et~al., 2018, \mn@doi [\mnras] {10.1093/mnras/sty1112}, \href
  {https://ui.adsabs.harvard.edu/abs/2018MNRAS.478.2034R} {478, 2034}

\bibitem[\protect\citeauthoryear{{Saifollahi}, {Trujillo}, {Beasley},
  {Peletier}  \& {Knapen}}{{Saifollahi} et~al.}{2021}]{Saifollahi2021}
{Saifollahi} T.,  {Trujillo} I.,  {Beasley} M.~A.,  {Peletier} R.~F.,
  {Knapen} J.~H.,  2021, \mn@doi [\mnras] {10.1093/mnras/staa3016}, \href
  {https://ui.adsabs.harvard.edu/abs/2021MNRAS.502.5921S} {502, 5921}

\bibitem[\protect\citeauthoryear{{Sakai}, {Madore}  \& {Freedman}}{{Sakai}
  et~al.}{1996}]{Sakai1996}
{Sakai} S.,  {Madore} B.~F.,   {Freedman} W.~L.,  1996, \mn@doi [\apj]
  {10.1086/177096}, \href
  {https://ui.adsabs.harvard.edu/abs/1996ApJ...461..713S} {461, 713}

\bibitem[\protect\citeauthoryear{{Salaris}, {Chieffi}  \&
  {Straniero}}{{Salaris} et~al.}{1993}]{Salaris1993}
{Salaris} M.,  {Chieffi} A.,   {Straniero} O.,  1993, \mn@doi [\apj]
  {10.1086/173105}, \href
  {https://ui.adsabs.harvard.edu/abs/1993ApJ...414..580S} {414, 580}

\bibitem[\protect\citeauthoryear{{Sales}, {Navarro}, {Pe{\~n}afiel}, {Peng},
  {Lim}  \& {Hernquist}}{{Sales} et~al.}{2020}]{Sales2020}
{Sales} L.~V.,  {Navarro} J.~F.,  {Pe{\~n}afiel} L.,  {Peng} E.~W.,  {Lim} S.,
   {Hernquist} L.,  2020, \mn@doi [\mnras] {10.1093/mnras/staa854}, \href
  {https://ui.adsabs.harvard.edu/abs/2020MNRAS.494.1848S} {494, 1848}

\bibitem[\protect\citeauthoryear{{Sandage}}{{Sandage}}{1976}]{Sandage1976}
{Sandage} A.,  1976, \mn@doi [\aj] {10.1086/111975}, \href
  {https://ui.adsabs.harvard.edu/abs/1976AJ.....81..954S} {81, 954}

\bibitem[\protect\citeauthoryear{{Sandage} \& {Binggeli}}{{Sandage} \&
  {Binggeli}}{1984}]{Sandage1984}
{Sandage} A.,  {Binggeli} B.,  1984, \mn@doi [\aj] {10.1086/113588}, \href
  {https://ui.adsabs.harvard.edu/abs/1984AJ.....89..919S} {89, 919}

\bibitem[\protect\citeauthoryear{{Schlafly} \& {Finkbeiner}}{{Schlafly} \&
  {Finkbeiner}}{2011}]{Schlafly2011}
{Schlafly} E.~F.,  {Finkbeiner} D.~P.,  2011, \mn@doi [\apj]
  {10.1088/0004-637X/737/2/103}, \href
  {https://ui.adsabs.harvard.edu/abs/2011ApJ...737..103S} {737, 103}

\bibitem[\protect\citeauthoryear{{Schlegel}, {Finkbeiner}  \&
  {Davis}}{{Schlegel} et~al.}{1998}]{Schlegel1998}
{Schlegel} D.~J.,  {Finkbeiner} D.~P.,   {Davis} M.,  1998, \mn@doi [\apj]
  {10.1086/305772}, \href
  {https://ui.adsabs.harvard.edu/abs/1998ApJ...500..525S} {500, 525}

\bibitem[\protect\citeauthoryear{{Sersic}}{{Sersic}}{1968}]{Sersic1968}
{Sersic} J.~L.,  1968, {Atlas de Galaxias Australes}

\bibitem[\protect\citeauthoryear{{Shen} et~al.,}{{Shen}
  et~al.}{2021}]{Shen2021}
{Shen} Z.,  et~al., 2021, \mn@doi [\apjl] {10.3847/2041-8213/ac0335}, \href
  {https://ui.adsabs.harvard.edu/abs/2021ApJ...914L..12S} {914, L12}

\bibitem[\protect\citeauthoryear{{Simon}, {Bolatto}, {Leroy}  \&
  {Blitz}}{{Simon} et~al.}{2003}]{Simon2003}
{Simon} J.~D.,  {Bolatto} A.~D.,  {Leroy} A.,   {Blitz} L.,  2003, \mn@doi
  [\apj] {10.1086/378200}, \href
  {https://ui.adsabs.harvard.edu/abs/2003ApJ...596..957S} {596, 957}

\bibitem[\protect\citeauthoryear{{Smercina} et~al.,}{{Smercina}
  et~al.}{2020}]{Smercina2020}
{Smercina} A.,  et~al., 2020, \mn@doi [\apj] {10.3847/1538-4357/abc485}, \href
  {https://ui.adsabs.harvard.edu/abs/2020ApJ...905...60S} {905, 60}

\bibitem[\protect\citeauthoryear{{Sorgho}, {Foster}, {Carignan}  \&
  {Chemin}}{{Sorgho} et~al.}{2019}]{Sorgho2019}
{Sorgho} A.,  {Foster} T.,  {Carignan} C.,   {Chemin} L.,  2019, \mn@doi
  [\mnras] {10.1093/mnras/stz696}, \href
  {https://ui.adsabs.harvard.edu/abs/2019MNRAS.486..504S} {486, 504}

\bibitem[\protect\citeauthoryear{{Spitler} \& {Forbes}}{{Spitler} \&
  {Forbes}}{2009}]{Spitler2009}
{Spitler} L.~R.,  {Forbes} D.~A.,  2009, \mn@doi [\mnras]
  {10.1111/j.1745-3933.2008.00567.x}, \href
  {https://ui.adsabs.harvard.edu/abs/2009MNRAS.392L...1S} {392, L1}

\bibitem[\protect\citeauthoryear{{Stone}, {Arora}, {Courteau}  \&
  {Cuillandre}}{{Stone} et~al.}{2021}]{Stone2021}
{Stone} C.~J.,  {Arora} N.,  {Courteau} S.,   {Cuillandre} J.-C.,  2021,
  \mn@doi [\mnras] {10.1093/mnras/stab2709}, \href
  {https://ui.adsabs.harvard.edu/abs/2021MNRAS.508.1870S} {508, 1870}

\bibitem[\protect\citeauthoryear{{Taylor} et~al.,}{{Taylor}
  et~al.}{2011}]{Taylor2011}
{Taylor} E.~N.,  et~al., 2011, \mn@doi [\mnras]
  {10.1111/j.1365-2966.2011.19536.x}, \href
  {https://ui.adsabs.harvard.edu/abs/2011MNRAS.418.1587T} {418, 1587}

\bibitem[\protect\citeauthoryear{{Toloba} et~al.,}{{Toloba}
  et~al.}{2016}]{Toloba2016}
{Toloba} E.,  et~al., 2016, \mn@doi [\apjl] {10.3847/2041-8205/816/1/L5}, \href
  {https://ui.adsabs.harvard.edu/abs/2016ApJ...816L...5T} {816, L5}

\bibitem[\protect\citeauthoryear{{Toloba} et~al.,}{{Toloba}
  et~al.}{2018}]{Toloba2018}
{Toloba} E.,  et~al., 2018, \mn@doi [\apjl] {10.3847/2041-8213/aab603}, \href
  {https://ui.adsabs.harvard.edu/abs/2018ApJ...856L..31T} {856, L31}

\bibitem[\protect\citeauthoryear{{Trujillo}, {Roman}, {Filho}  \& {S{\'a}nchez
  Almeida}}{{Trujillo} et~al.}{2017}]{Trujillo2017}
{Trujillo} I.,  {Roman} J.,  {Filho} M.,   {S{\'a}nchez Almeida} J.,  2017,
  \mn@doi [\apj] {10.3847/1538-4357/aa5cbb}, \href
  {https://ui.adsabs.harvard.edu/abs/2017ApJ...836..191T} {836, 191}

\bibitem[\protect\citeauthoryear{Virtanen et~al.,}{Virtanen
  et~al.}{2020}]{Scipy2020}
Virtanen P.,  et~al., 2020, \mn@doi [Nature Methods]
  {10.1038/s41592-019-0686-2}, \href {https://rdcu.be/b08Wh} {17, 261}

\bibitem[\protect\citeauthoryear{{Willmer}}{{Willmer}}{2018}]{Willmer2018}
{Willmer} C. N.~A.,  2018, \mn@doi [\apjs] {10.3847/1538-4365/aabfdf}, \href
  {https://ui.adsabs.harvard.edu/abs/2018ApJS..236...47W} {236, 47}

\bibitem[\protect\citeauthoryear{{Yagi}, {Koda}, {Komiyama}  \&
  {Yamanoi}}{{Yagi} et~al.}{2016}]{Yagi2016}
{Yagi} M.,  {Koda} J.,  {Komiyama} Y.,   {Yamanoi} H.,  2016, \mn@doi [\apjs]
  {10.3847/0067-0049/225/1/11}, \href
  {https://ui.adsabs.harvard.edu/abs/2016ApJS..225...11Y} {225, 11}

\bibitem[\protect\citeauthoryear{{Yun}, {Ho}  \& {Lo}}{{Yun}
  et~al.}{1994}]{Yun1994}
{Yun} M.~S.,  {Ho} P.~T.~P.,   {Lo} K.~Y.,  1994, \mn@doi [\nat]
  {10.1038/372530a0}, \href
  {https://ui.adsabs.harvard.edu/abs/1994Natur.372..530Y} {372, 530}

\bibitem[\protect\citeauthoryear{{de Blok} et~al.,}{{de Blok}
  et~al.}{2018}]{deBlok2018}
{de Blok} W.~J.~G.,  et~al., 2018, \mn@doi [\apj] {10.3847/1538-4357/aad557},
  \href {https://ui.adsabs.harvard.edu/abs/2018ApJ...865...26D} {865, 26}

\bibitem[\protect\citeauthoryear{{van Dokkum}, {Abraham}, {Merritt}, {Zhang},
  {Geha}  \& {Conroy}}{{van Dokkum} et~al.}{2015}]{vanDokkum2015}
{van Dokkum} P.~G.,  {Abraham} R.,  {Merritt} A.,  {Zhang} J.,  {Geha} M.,
  {Conroy} C.,  2015, \mn@doi [\apjl] {10.1088/2041-8205/798/2/L45}, \href
  {https://ui.adsabs.harvard.edu/abs/2015ApJ...798L..45V} {798, L45}

\bibitem[\protect\citeauthoryear{{van Dokkum} et~al.,}{{van Dokkum}
  et~al.}{2017}]{vanDokkum2017}
{van Dokkum} P.,  et~al., 2017, \mn@doi [\apjl] {10.3847/2041-8213/aa7ca2},
  \href {https://ui.adsabs.harvard.edu/abs/2017ApJ...844L..11V} {844, L11}

\bibitem[\protect\citeauthoryear{{van Dokkum} et~al.,}{{van Dokkum}
  et~al.}{2018}]{vDokkum2018}
{van Dokkum} P.,  et~al., 2018, \mn@doi [\nat] {10.1038/nature25767}, \href
  {https://ui.adsabs.harvard.edu/abs/2018Natur.555..629V} {555, 629}

\bibitem[\protect\citeauthoryear{{van Dokkum}, {Danieli}, {Abraham}, {Conroy}
  \& {Romanowsky}}{{van Dokkum} et~al.}{2019}]{vDokkum2019}
{van Dokkum} P.,  {Danieli} S.,  {Abraham} R.,  {Conroy} C.,   {Romanowsky}
  A.~J.,  2019, \mn@doi [\apjl] {10.3847/2041-8213/ab0d92}, \href
  {https://ui.adsabs.harvard.edu/abs/2019ApJ...874L...5V} {874, L5}

\makeatother
\end{thebibliography}



\appendix

\section{Photometric Transformations between the HST and HSC Systems} \label{sec:transformation}

 In order to use HST observations to derive the completeness of our HSC observations, we needed to convert photometry between the two filter systems. 
 Due to the lack of HST to HSC photometric transformations in the literature, we derived our own equations. Following the procedure outlined in \cite{Komiyama2018}, we convolved the HSC and HST filter transmission curves of interest with stellar spectra from the \cite{Gunn1983} and \cite{Pickles1998} spectral atlases, which together contain stars with a broad range of colour. Only the HST F606W and F814W to HSC {\it g} and {\it i} band filters are needed for our study.  The integrated fluxes in the bandpasses were converted to either AB magnitudes (HSC filters) or VEGA magnitudes (HST filters).

\begin{figure}
	\includegraphics[width=\columnwidth]{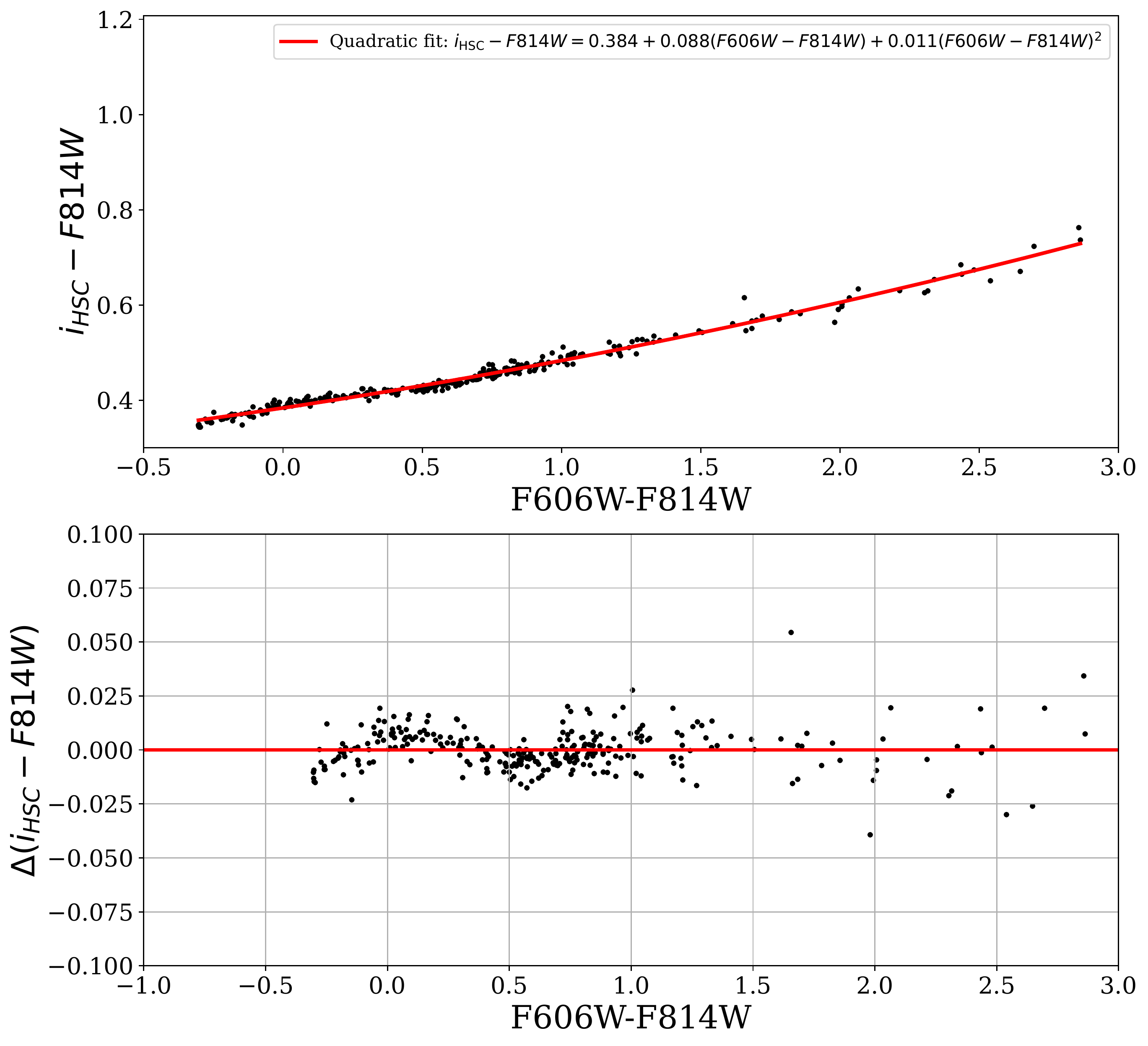}
    \caption{Conversion from the HST F814W (Vega) to $i\mathrm{_{HSC}}$ (AB) filter systems using a quadratic fit to the F606W$-$F814W colour.  The top panel shows the magnitude difference of the atlas stars as black points while the bottom panel shows the residuals of the same stars from the quadratic fit.}
    \label{fig:i_814}
\end{figure}

\begin{figure}
	\includegraphics[width=\columnwidth]{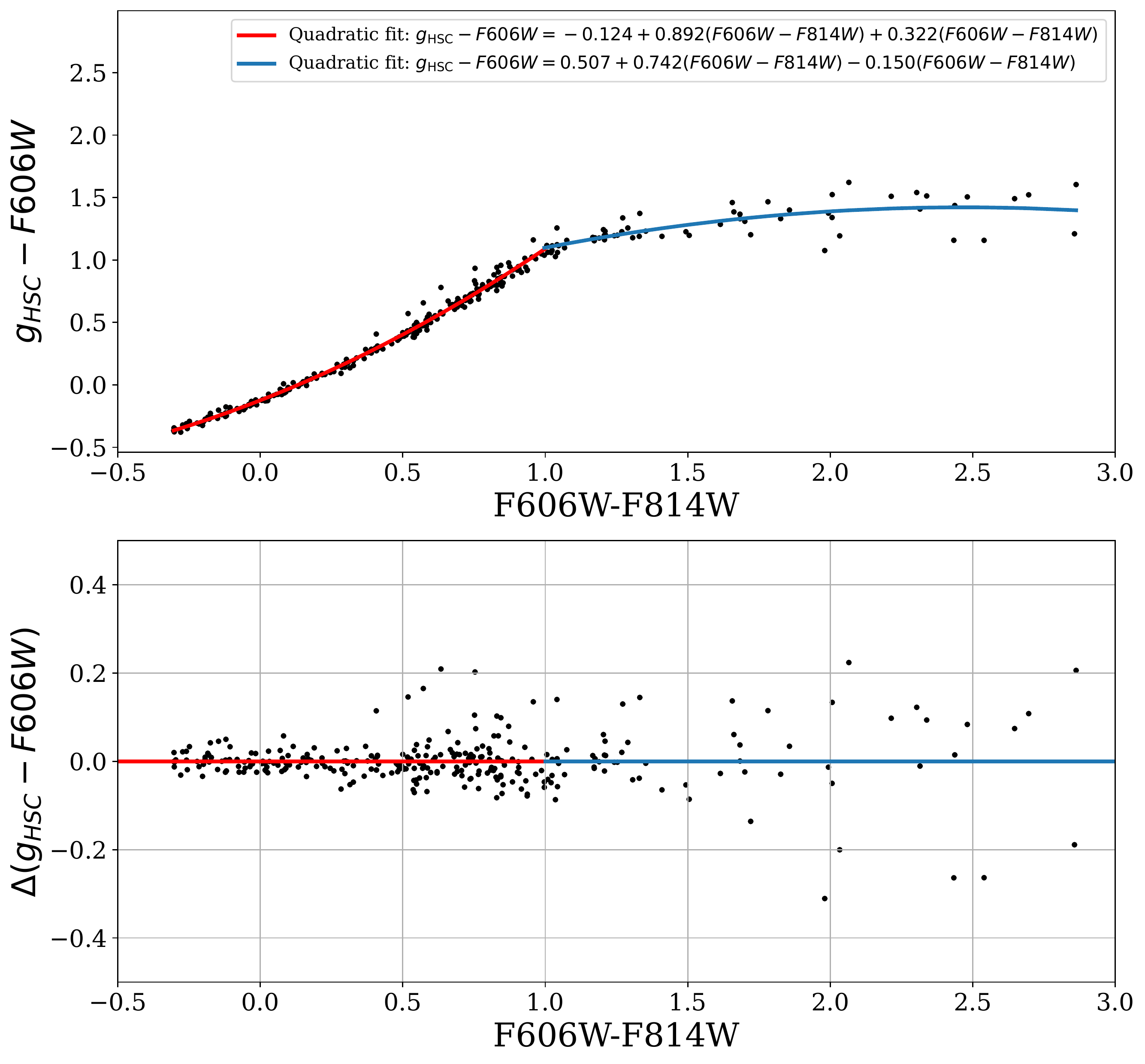}
    \caption{Conversion from the HST F606W (Vega) to $g\mathrm{_{HSC}}$ (AB) filter systems using a quadratic fit to the F606W$-$F814W. Divergent behaviour is seen beyond F606W$-$F814W$=$1.0, thus requiring a separate quadratic fit in this region. The top panel shows the magnitude difference of the atlas stars as black points while the bottom panel shows the residuals of the same stars from the quadratic fit.}
    \label{fig:g_606}
\end{figure}

 Figures \ref{fig:i_814} and \ref{fig:g_606} show the difference in derived magnitudes between the two filter systems as a function the F606W$-$F814W colour. A quadratic polynomial fit was found to be sufficient to describe the conversion between the two red bands: 

\begin{multline}
    i_{\rm HSC} - F814W = 0.384 + 0.088 (F606W-F814W) \\
    + 0.011 (F606W-F814W)^2
    \end{multline}

The difference between the blue bands exhibits a break at $\mathrm{F606W}-\mathrm{F814W} \sim 1$, requiring two separate quadratic polynomial fits:

\begin{multline}
    g_{\rm HSC} - F606W  = -0.124 + 0.892 (F606W-F814W) \\
    +0.322 (F606W-F814W)^2
\end{multline}

\noindent for $\mathrm{F606W}-\mathrm{F814W}<1.0$, and 

\begin{multline}
    g_{\rm HSC} - F606W = 0.507 + 0.742 (F606W-F814W) \\
    -0.15 (F606W-F814W)^2
    \end{multline}
    
\noindent for $\mathrm{F606W}-\mathrm{F814W}>1.0$. In all cases, the HSC magnitudes are on the AB system and the HST magnitudes are on the Vega system.  

The {\it i}-band fit has a very low scatter -- the residuals are typically less than 0.025 mag even at the reddest colours. While there is some higher-order behaviour at colours bluer than F606W$-$F814W $=$ 0.75 which our polynomial fit fails to capture, this level of precision is not required for our purposes.  On the other hand, the {\it g}-band fit exhibits higher residuals at all colours, ranging from $\sim 0.075$ mag at colours bluer than F606W$-$F814W $=$ 1.0 to $\sim 0.2$ mag beyond this limit. This is not unexpected given that $g\mathrm{_{HSC}}$ is considerably bluer in transmission than F606W. Nonetheless, this description is adequate for our study.


\bsp	
\label{lastpage}
\end{document}